\documentclass[fleqn]{llncs}
\usepackage{tcmd}
\usepackage{version}
\usepackage{pstricks}
\usepackage{pst-node}

\title{A Thread Calculus with Molecular Dynamics%
       \thanks{This research has been partly carried out in the
               framework of the GLANCE-project MICROGRIDS, which is
               funded by the Netherlands Organisation for Scientific
               Research (NWO).}}

\author{J.A. Bergstra \and C.A. Middelburg%
        \thanks{The work presented in this paper has been partly carried
                out while the first author was also at Utrecht
                University, Department of Philosophy, and the second
                author was also at Eindhoven University of Technology,
                Department of Mathematics and Computer Science.}}
\institute{Programming Research Group,
           University of Amsterdam, \\
           Kruislaan~403, 1098~SJ~Amsterdam, the Netherlands \\
           \email{J.A.Bergstra@uva.nl,C.A.Middelburg@uva.nl}
           }

\begin{document}
\maketitle

\begin{abstract}
We present a theory of threads, interleaving of threads, and interaction
between threads and services with features of molecular dynamics, a
model of computation that bears on computations in which dynamic data
structures are involved.
Threads can interact with services of which the states consist of
structured data objects and computations take place by means of actions
which may change the structure of the data objects.
The features introduced include restriction of the scope of names used
in threads to refer to data objects.
Because that feature makes it troublesome to provide a model based on
structural operational semantics and bisimulation, we construct a
projective limit model for the theory.
\\[1.5ex]
{\sl Keywords:}
thread calculus, thread algebra, molecular dynamics, restriction,
projective limit model.
\\[1.5ex]
{\sl 1998 ACM Computing Classification:}
D.1.3, D.1.5, D.3.3, F.1.1, F.1.2, F.3.2.
\end{abstract}

\section{Introduction}
\label{sect-intro}

A thread is the behaviour of a deterministic sequential program under
execution.
Multi-threading refers to the concurrent existence of several threads
in a program under execution.
Multi-threading is the dominant form of concurrency provided by
contemporary programming languages such as Java~\cite{GJSB00a} and
C\#~\cite{HWG03a}.
We take the line that arbitrary interleaving, on which theories and
models about concurrent processes such as \ACP~\cite{BK84b}, the
$\pi$-calculus~\cite{MPW92c} and the Actor model~\cite{AMST97a} are
based, is not the most appropriate abstraction when dealing with
multi-threading.
In the case of multi-threading, more often than not some deterministic
interleaving strategy is used.
In~\cite{BM04c}, we introduced a number of plausible deterministic
interleaving strategies for multi-threading.
We proposed to use the phrase strategic interleaving for the more
constrained form of interleaving obtained by using such a strategy.
We also introduced a feature for interaction of threads with services.
The algebraic theory of threads, multi-threading, and interaction
of threads with services is called thread algebra.

In the current paper, we extend thread algebra with features of
molecular dynamics, a model of computation that bears on computations in
which dynamic data structures are involved.
Threads can interact with services of which the states consist of
structured data objects and computations take place by means of actions
which may change the structure of the data objects.
The states resemble collections of molecules composed of atoms and the
actions can be considered to change the structure of molecules like in
chemical reactions.
We elaborate on the model described informally in~\cite{BB02a}.
The additional features include a feature to restrict the scope of
names used in threads to refer to data objects.
That feature turns thread algebra into a calculus.
Although it occurs in quite another setting, it is reminiscent of
restriction in the $\pi$-calculus~\cite{MPW92c}.

In thread algebra, we abandon the point of view that arbitrary
interleaving is the most appropriate abstraction when dealing with
multi-threading.
The following points illustrate why we find difficulty in taking that
point of view:
(a)~whether the interleaving of certain threads leads to deadlock
depends on the interleaving strategy used;
(b)~sometimes deadlock takes place with a particular interleaving
strategy whereas arbitrary interleaving would not lead to deadlock, and
vice versa.
Demonstrations of (a) and (b) are given in~\cite{BM04c}
and~\cite{BM05c}, respectively.
Arbitrary interleaving and interleaving according to some deterministic
interleaving strategy are two extreme forms of interleaving, but
nevertheless they are both abstractions for multi-threading.
Even in the case where real multi-threading is interleaving according to
an interleaving strategy with some non-deterministic aspects, there is
no reason to simply assume that arbitrary interleaving is the most
adequate abstraction.

The thread-service dichotomy that we make in thread algebra is useful
for the following reasons:
(a)~for services, a state-based description is generally more convenient
than an action-based description whereas it is the other way round for
threads;
(b)~the interaction between threads and services is of an asymmetric
nature.
Evidence of both (a) and (b) is produced in~\cite{BM05c} by the
established connections of threads and services with processes as
considered in an extension of \ACP\ with conditions introduced
in~\cite{BM05a}.

We started the work on thread algebra with the object to develop a
theory about threads, multi-threading and interaction of threads with
services that is useful for (a)~gaining insight into the semantic issues
concerning the multi-threading related features found in contemporary
programming languages, and (b)~simplified formal description and
analysis of programs in which multi-threading is involved.

Although thread algebra is concerned with the constrained form of
interleaving found in multi-threading as provided by contemporary
programming languages, not all relevant details of multi-threading as
provided by those languages can be modelled with thread algebra.
The details concerned come up where multi-threading is adjusted to
the object-orientation of those languages.
It gives rise to a form of thread forking where thread forking is
divided into creating a thread object and starting the execution of the
thread associated with the created object.
Setting up a framework in which these details can be modelled as well is
the main objective with which we have extended thread algebra with
features of molecular dynamics.
The form of thread forking mentioned above is modelled in this paper
using the thread calculus developed.
The feature to restrict the scope of names used in threads to refer to
data objects turns out to be indispensable when modelling this form of
thread forking.

The construction of a model for the full thread calculus developed in
this paper by means of a structural operational semantics and an
appropriate version of bisimilarity is troublesome.
This is mainly due to the feature to restrict the scope of names used in
threads to refer to data objects.
In fact, this feature complicates matters to such an extent that the
structural operational semantics would add only marginally to a better
understanding and the appropriate version of bisimilarity would be
difficult to comprehend.
Therefore, we provide instead a projective limit model.
In process algebra, a projective limit model has been given for the
first time in~\cite{BK84b}.
Following~\cite{Kra87a}, we make the domain of the projective limit
model into a metric space to show, using Banach's fixed point theorem,
that operations satisfying a guardedness condition have unique fixed
points.
Metric spaces have also been applied by others in concurrency theory,
either to establish uniqueness results for recursion
equations~\cite{BBKM84a} or to solve domain equations for process
domains~\cite{BZ82a}.
We also determine the position in the arithmetical hierarchy of the
equality relation in the projective limit model.

Thread forking is inherent in multi-threading.
However, we will not introduce thread interleaving and thread forking
combined.
Thread forking is presented at a later stage as an extension.
This is for expository reasons only.
The formulations of many results, as well as their proofs, would be
complicated by introducing thread forking at an early stage because the
presence of thread forking would be accountable to many exceptions in
the results.
In the set-up in which thread forking is introduced later on, we can
simply summarize which results need to be adapted to the presence of
thread forking and how.

Thread algebra is a design on top of an algebraic theory of the
behaviour of deterministic sequential programs under execution
introduced in~\cite{BL02a} under the name basic polarized process
algebra.
Prompted by the development of thread al\-gebra, basic polarized process
algebra has been renamed to basic thread algebra.

Dynamic data structures modelled using molecular dynamics can
straightforwardly be implemented in programming languages ranging from
PASCAL~\cite{Wir71a} to C\#~\cite{HWG03a} through pointers or
references, provided that fields are not added or removed dynamically.
Using molecular dynamics, we need not be aware of the existence of the
pointers used for linking data.
The name molecular dynamics refers to the molecule metaphor used above.
By that, there is no clue in the name itself to what it stands for.
Remedying this defect, the recent upgrade of molecular dynamics
presented in~\cite{BM08d} is called data linkage dynamics.
Chemical abstract machines~\cite{BB92d} are also explained using a
molecule metaphor.
However, molecular dynamics is concerned with the structure of
molecule-resembling data, whereas chemical abstract machines are
concerned with reaction between molecule-resembling processes.

We can summarize the main contributions of this paper as follows:
\begin{enumerate}
\item
the extension of thread algebra with features of molecular dynamics,
including operators to restrict the scope of names used in molecular
dynamics;
\item
the modelling of the form of thread forking found in contemporary
programming languages such as Java and C\# in the resulting thread
calculus;
\item
the construction of a projective limit model for the thread calculus;
\item
the result that equality in the projective limit model is a
$\rPi^0_1$-relation.
\end{enumerate}

The body of this paper consists of two parts.
The first part (Sections~\ref{sect-BTA}--\ref{sect-TCf-Java}) is
concerned with the thread calculus in itself.
To bring structure in the thread calculus, it is presented in a modular
way.
The second part (Sections~\ref{sect-projlim-TC}--\ref{sect-PGLDf-eq}) is
concerned with the projective limit model for the thread calculus.

The first part is organized as follows.
First, we review basic thread algebra (Section~\ref{sect-BTA}).
Then, we extend basic thread algebra to a theory of threads,
interleaving of threads and interaction of threads with services
(Sections~\ref{sect-TA} and~\ref{sect-TAtsc}), and introduce
recursion in this setting (Section~\ref{sect-REC}).
Next, we propose a state-based approach to describe services
(Section~\ref{sect-service-descr}) and use it to describe services for
molecular dynamics (Section~\ref{sect-MDS}).
After that, we introduce a feature to restrict the scope of names used
in threads to refer to data objects (Section~\ref{sect-TC}).
Following this, we introduce the approximation induction principle to
reason about infinite threads (Section~\ref{sect-AIP}).
Finally, we introduce a basic form of thread forking
(Section~\ref{sect-TCf}) and illustrate how the restriction feature can
be used to model a form of thread forking found in contemporary
programming languages (Section~\ref{sect-TCf-Java}).

The second part is organized as follows.
First, we construct the projective limit model for the thread calculus
without thread forking in two steps
(Sections~\ref{sect-projlim-TC}, \ref{sect-projlim-metric},
and~\ref{sect-projlim-REC}).
Then, we show that recursion equations satisfying a guardedness
condition have unique solutions in this model
(Section~\ref{sect-guarded-rec-eqns}).
Next, we determine the position in the arithmetical hierarchy of the
equality relation in this model (Section~\ref{sect-projlim-eq}).
After that, we outline the adaptation of the projective limit model to
thread forking (Section~\ref{sect-projlim-TCf}) and dwell briefly on the
behavioural equivalence of programs from a simple program notation with
support of thread forking in the resulting model
(Section~\ref{sect-PGLDf-eq}).

The proofs of the theorems and propositions for which no proof is given
in this paper can be found in~\cite{BM06a}.
In Sections~\ref{sect-projlim-metric}--\ref{sect-guarded-rec-eqns},
some familiarity with metric spaces is assumed.
The definitions of all notions concerning metric spaces that are assumed
known in those sections can be found in most introductory textbooks on
topology.
We mention~\cite{Cro89a} as an example of an introductory textbook in
which those notions are introduced in an intuitively appealing way.

\section{Basic Thread Algebra}
\label{sect-BTA}

In this section, we review \BTA\ (Basic Thread Algebra), introduced
in~\cite{BL02a} under the name \BPPA\ (Basic Polarized Process Algebra).
\BTA\ is a form of process algebra which is tailored to the description
of the behaviour of deterministic sequential programs under execution.

In \BTA, it is assumed that a fixed but arbitrary set $\BAct$ of
\emph{basic actions}, with $\Tau \notin \BAct$, has been given.
Besides, $\Tau$ is a special basic action.
We write $\BActTau$ for $\BAct \union \set{\Tau}$.
A thread performs basic actions in a sequential fashion.
Upon each basic action performed, a reply from the execution environment
of the thread determines how it proceeds.
The possible replies are $\True$ and $\False$.
Performing $\Tau$, which is considered performing an internal action,
always leads to the reply $\True$.

The signature of \BTA\ consists of the following constants and
operators:
\begin{iteml}
\item
the \emph{deadlock} constant $\Dead$;
\item
the \emph{termination} constant $\Stop$;
\item
for each $a \in \BActTau$, a binary \emph{postconditional
composition} operator $\pcc{\ph}{a}{\ph}\,$.
\end{iteml}
Throughout the paper, we assume that there is a countably infinite set
of variables, including $x,y,z,x_1,x'_1,x_2,x'_2,\ldots {}$.
Terms over the signature of \BTA\ are built as usual
(see e.g.~\cite{ST99a,Wir90a}).
Terms that contain no variables are called closed terms.
We use infix notation for postconditional composition.
We introduce \emph{action prefixing} as an abbreviation:
$a \bapf p$, where $p$ is a term over the signature of \BTA, abbreviates
$\pcc{p}{a}{p}$.

The thread denoted by a closed term of the form $\pcc{p}{a}{q}$ will
first perform $a$, and then proceed as the thread denoted by $p$
if the reply from the execution environment is $\True$ and proceed as
the thread denoted by $q$ if the reply from the execution environment is
$\False$.
The threads denoted by $\Dead$ and $\Stop$ will become inactive and
terminate, respectively.

\BTA\ has only one axiom.
This axiom is given in Table~\ref{axioms-BTA}.%
\begin{table}[!tb]
\caption{Axiom of \BTA}
\label{axioms-BTA}
\begin{eqntbl}
\begin{axcol}
\pcc{x}{\Tau}{y} = \pcc{x}{\Tau}{x}                & \axiom{T1}
\end{axcol}
\end{eqntbl}
\end{table}
Using the abbreviation introduced above, axiom T1 can be written as
follows: $\pcc{x}{\Tau}{y} = \Tau \bapf x$.

Henceforth, we will write $\BTA(A)$ for \BTA\ with the set of basic
actions $\BAct$ fixed to be the set $A$.

As mentioned above, the behaviour of a thread depends upon its execution
environment.
Each basic action performed by the thread is taken as a command to be
processed by the execution environment.
At any stage, the commands that the execution environment can accept
depend only on its history, i.e.\ the sequence of commands processed
before and the sequence of replies produced for those commands.
When the execution environment accepts a command, it will produce a
reply value.
Whether the reply is $\True$ or $\False$ usually depends on the
execution history.
However, it may also depend on external conditions.
For example, when the execution environment accepts a command to write a
file to a memory card, it will usually produce a positive reply, but not
if the memory card turns out to be write-protected.

In the structural operational semantics of \BTA, we represent an
execution environment by a function
$\rho: \seqof{(\BAct \x \set{\True,\False})} \to
       \setof{(\BAct \x \set{\True,\False})}$
that satisfies the following condition:
$\tup{a,b} \not\in \rho(\alpha) \Implies
 \rho(\alpha \conc \seq{\tup{a,b}}) = \emptyset$
for all $a \in \BAct$, $b \in \set{\True,\False}$ and
$\alpha \in \seqof{(\BAct \x \set{\True,\False})}$.%
\footnote
{We write $\seqof{D}$ for the set of all finite sequences with elements
 from set $D$, and $\neseqof{D}$ for the set of all non-empty finite
 sequences with elements from set $D$.
 We write
 $\emptyseq$ for the empty sequence,
 $\seq{d}$ for the sequence having $d$ as sole element, and
 $\alpha \conc \beta$ for the concatenation of finite sequences $\alpha$
 and $\beta$.
 We assume the usual laws for concatenation of finite sequences.
}
We write $\ExEnv$ for the set of all those functions.
Given an execution environment $\rho \in \ExEnv$ and a basic action
$a \in \BAct$,
the \emph{derived} execution environment of $\rho$ after processing $a$
with a \emph{positive} reply, written $\smash{\derivep{a} \rho}$, is
defined by
$\smash{\derivep{a} \rho(\alpha) =
 \rho(\seq{\tup{a,\True}} \conc \alpha)}$;
and
the \emph{derived} execution environment of $\rho$ after processing $a$
with a \emph{negative} reply, written $\smash{\derivem{a} \rho}$, is
defined by
$\smash{\derivem{a} \rho(\alpha) =
 \rho(\seq{\tup{a,\False}} \conc \alpha)}$.

The following transition relations on closed terms over the signature of
\BTA\ are used in the structural operational semantics of \BTA:
\begin{iteml}
\item
a binary relation $\astep{\cfg{\ph,\rho}}{a}{\cfg{\ph,\rho'}}$ for each
$a \in \BActTau$ and $\rho,\rho' \in \ExEnv$;
\item
a unary relation $\sterm{\ph}$;
\item
a unary relation $\dterm{\ph}$;
\item
a unary relation $\sdterm{\ph}$.
\end{iteml}
The four kinds of transition relations are called the
\emph{action step}, \emph{termination}, \emph{deadlock}, and
\emph{termination or deadlock} relations, respectively.
They can be explained as follows:
\begin{iteml}
\item
$\astep{\cfg{p,\rho}}{a}{\cfg{p',\rho'}}$:
in execution environment $\rho$, thread $p$ can perform action $a$ and
after that proceed as thread $p'$ in execution environment $\rho'$;
\item
$\sterm{p}$:
thread $p$ cannot but terminate successfully;
\item
$\dterm{p}$:
thread $p$ cannot but become inactive;
\item
$\sdterm{p}$:
thread $p$ cannot but terminate successfully or become inactive.
\end{iteml}
The termination or deadlock relation is an auxiliary relation needed
when we extend \BTA\ in Section~\ref{sect-TA}.

The structural operational semantics of \BTA\ is described by the
transition rules given in Table~\ref{sos-BTA}.%
\begin{table}[!tb]
\caption{Transition rules of \BTA}
\label{sos-BTA}
\begin{ruletbl}
\Rule
{\phantom{\sterm{\Stop}}}
{\sterm{\Stop}}
\qquad
\Rule
{\phantom{\sterm{\Stop}}}
{\dterm{\Dead}}
\hfill
\RuleC
{\phantom{\sterm{\Stop}}}
{\astep{\cfg{\pcc{x}{\Tau}{y},\rho}}{\Tau}{\cfg{x,\rho}}}
{\;\phantom{\tup{a,\False} \in \rho(\emptyseq)}}
\\
\RuleC
{\phantom{\sterm{\Stop}}}
{\astep{\cfg{\pcc{x}{a}{y},\rho}}{a}{\cfg{x,\derivep{a} \rho}}}
{\tup{a,\True} \in \rho(\emptyseq)}
\qquad
\RuleC
{\phantom{\sterm{\Stop}}}
{\astep{\cfg{\pcc{x}{a}{y},\rho}}{a}{\cfg{y,\derivem{a} \rho}}}
{\tup{a,\False} \in \rho(\emptyseq)}
\\
\Rule
{\sterm{x}}
{\sdterm{x}}
\qquad
\Rule
{\dterm{x}}
{\sdterm{x}}
\end{ruletbl}
\end{table}
In this table $a$ stands for an arbitrary action from $\BAct$.

Bisimulation equivalence is defined as follows.
A \emph{bisimulation} is a symmetric binary relation $B$ on closed terms
over the signature of \BTA\ such that for all closed terms $p$ and $q$:
\begin{iteml}
\item
if $B(p,q)$ and $\astep{\cfg{p,\rho}}{a}{\cfg{p',\rho'}}$, then
there is a $q'$ such that
$\astep{\cfg{q,\rho}}{a}{\cfg{q',\rho'}}$ and $B(p',q')$;
\item
if $B(p,q)$ and $\sterm{p}$, then $\sterm{q}$;
\item
if $B(p,q)$ and $\dterm{p}$, then $\dterm{q}$.
\end{iteml}
Two closed terms $p$ and $q$ are \emph{bisimulation equivalent},
written $p \bisim q$, if there exists a bisimulation $B$ such that
$B(p,q)$.

Bisimulation equivalence is a congruence with respect to the
postconditional composition operators.
This follows immediately from the fact that the transition rules for
these operators are in the path format (see e.g.~\cite{AFV00a}).
The axiom given in Table~\ref{axioms-BTA} is sound with respect to
bisimulation equivalence.

\section{Strategic Interleaving of Threads}
\label{sect-TA}

In this section, we take up the extension of \BTA\ to a theory about
threads and multi-threading by introducing a very simple interleaving
strategy.
This interleaving strategy, as various other plausible interleaving
strategies, was first formalized in an extension of \BTA\
in~\cite{BM04c}.

It is assumed that the collection of threads to be interleaved takes
the form of a sequence of threads, called a \emph{thread vector}.
Strategic interleaving operators turn a thread vector of arbitrary
length into a single thread.
This single thread obtained via a strategic interleaving operator is
also called a multi-thread.
Formally, however multi-threads are threads as well.

The very simple interleaving strategy that we introduce here is called
\emph{cyclic interleaving}.%
\footnote
{Implementations of the cyclic interleaving strategy are usually called
 round-robin schedulers.}
Cyclic interleaving basically operates as follows: at each stage of the
interleaving, the first thread in the thread vector gets a turn to
perform a basic action and then the thread vector undergoes cyclic
permutation.
We mean by cyclic permutation of a thread vector that the first thread
in the thread vector becomes the last one and all others move one
position to the left.
If one thread in the thread vector deadlocks, the whole does not
deadlock till all others have terminated or deadlocked.
An important property of cyclic interleaving is that it is fair, i.e.\
there will always come a next turn for all active threads.
Other plausible interleaving strategies are treated in~\cite{BM04c}.
They can also be adapted to the features of molecular dynamics that will
be introduced in the current\linebreak[2] paper.

In order to extend \BTA\ to a theory about threads and multi-threading,
we introduce the unary operator $\csiop$.
This operator is called the strategic interleaving operator for cyclic
interleaving.
The thread denoted by a closed term of the form $\csi{\alpha}$ is the
thread that results from cyclic interleaving of the threads in the
thread vector denoted by $\alpha$.

The axioms for cyclic interleaving are given in
Table~\ref{axioms-cyclstrint}.%
\begin{table}[!tb]
\caption{Axioms for cyclic interleaving}
\label{axioms-cyclstrint}
\begin{eqntbl}
\begin{axcol}
\csi{\emptyseq} = \Stop                        & \axiom{CSI1}\\
\csi{\seq{\Stop}\conc \alpha} =
                                  \csi{\alpha} & \axiom{CSI2} \\
\csi{\seq{\Dead} \conc \alpha} =
                     \std{\csi{\alpha}} & \axiom{CSI3} \\
\csi{\seq{\Tau \bapf x}\conc \alpha} =
         \Tau \bapf \csi{\alpha \conc \seq{x}} & \axiom{CSI4} \\
\csi{\seq{\pcc{x}{a}{y}}\conc \alpha} =
           \pcc{\csi{\alpha \conc \seq{x}}}{a}
                  {\csi{\alpha \conc \seq{y}}} & \axiom{CSI5}
\end{axcol}
\end{eqntbl}
\end{table}
In CSI3, the auxiliary \emph{deadlock at termination} operator $\stdop$
is used to express that in the event of deadlock of one thread in the
thread vector, the whole deadlocks only after all others have terminated
or deadlocked.
The thread denoted by a closed term of the form $\std{p}$ is the thread
that results from turning termination into deadlock in the thread
denoted by $p$.
The axioms for deadlock at termination appear in
Table~\ref{axioms-StopToDead}.%
\begin{table}[!tb]
\caption{Axioms for deadlock at termination}
\label{axioms-StopToDead}
\begin{eqntbl}
\begin{axcol}
\std{\Stop} = \Dead                         & \axiom{S2D1} \\
\std{\Dead} = \Dead                      & \axiom{S2D2} \\
\std{\Tau \bapf x} = \Tau \bapf \std{x} & \axiom{S2D3} \\
\std{\pcc{x}{a}{y}} =
         \pcc{\std{x}}{a}{\std{y}} & \axiom{S2D4}
\end{axcol}
\end{eqntbl}
\end{table}
In Tables~\ref{axioms-cyclstrint} and~\ref{axioms-StopToDead}, $a$
stands for an arbitrary action from $\BAct$.

Henceforth, we will write \TA\ for \BTA\ extended with the strategic
interleaving operator for cyclic interleaving, the deadlock at
termination operator, and the axioms from Tables~\ref{axioms-cyclstrint}
and~\ref{axioms-StopToDead}, and we will write $\TA(A)$ for \TA\ with
the set of basic actions $\BAct$ fixed to be the set $A$.

\begin{example}
\label{example-TA}
The following equation is easily derivable from the axioms of \TA:
\begin{ldispl}
\hspace*{-2.5em}
\csi
 {\seq{\pcc{(a'_1 \bapf \Stop)}{a_1}{(a''_1 \bapf \Stop)}} \conc
  \seq{\pcc{(a'_2 \bapf \Stop)}{a_2}{(a''_2 \bapf \Stop)}}}
\\ \hspace*{-2.5em} \; {} =
\pcc{(\pcc{(a'_1 \bapf a'_2 \bapf \Stop)}{a_2}
          {(a'_1 \bapf a''_2 \bapf \Stop)})}
    {a_1}
    {
     (\pcc{(a''_1 \bapf a'_2 \bapf \Stop)}{a_2}
          {(a''_1 \bapf a''_2 \bapf \Stop)})}\;.
\end{ldispl}%
This equation shows clearly that the threads denoted by
$\pcc{(a'_1 \bapf \Stop)}{a_1}{(a''_1 \bapf \Stop)}$
and
$\pcc{(a'_2 \bapf \Stop)}{a_2}{(a''_2 \bapf \Stop)}$
are interleaved in a cyclic manner:
first the first thread performs $a_1$,
next the second thread performs $a_2$,
next the first thread performs $a'_1$ or $a''_1$
depending upon the reply on $a_1$,
next the second thread performs $a'_2$ or $a''_2$
depending upon the reply on $a_2$.
\end{example}

We can prove that each closed term over the signature of \TA\ can be
reduced to a closed term over the signature of \BTA.
\begin{theorem}[Elimination]
\label{thm-elim-TA}
For all closed terms $p$ over the signature of \TA, there exists a
closed term $q$ over the signature of \BTA\ such that $p = q$ is
derivable from the axioms of \TA.
\end{theorem}

The following proposition, concerning the cyclic interleaving of a
thread vector of length $1$, is easily proved using
Theorem~\ref{thm-elim-TA}.
\begin{proposition}
\label{prop-csi}
For all closed terms $p$ over the signature of \TA, the equation
$\csi{\seq{p}} = p$ is derivable from the axioms of \TA.
\end{proposition}
The equation $\csi{\seq{p}} = p$ from Proposition~\ref{prop-csi}
expresses the obvious fact that in the cyclic interleaving of a thread
vector of length $1$ no proper interleaving is involved.

The following are useful properties of the deadlock at termination
operator which are proved using Theorem~\ref{thm-elim-TA} as well.
\begin{proposition}
\label{prop-s2d-csi}
For all closed terms $p_1,\ldots,p_n$ over the signature of \TA, the
following equations are derivable from the axioms of \TA:
\begin{eqnarray}
& &
\std{\csi{\seq{p_1} \conc \ldots \conc \seq{p_n}}} =
\csi{\seq{\std{p_1}} \conc \ldots \conc
            \seq{\std{p_n}}}\;,
\label{s2d-1}
\\
& &
\std{\std{p_1}} = \std{p_1}\;.
\label{s2d-2}
\end{eqnarray}
\end{proposition}

The structural operational semantics of \TA\ is described by the
transition rules given in Tables~\ref{sos-BTA}
and~\ref{sos-cyclstrint}.
\begin{table}[!tb]
\caption{Transition rules for cyclic interleaving and
  deadlock at termination}
\label{sos-cyclstrint}
\begin{ruletbl}
\Rule
{\sterm{x_1},\ldots,\sterm{x_k},
 \astep{\cfg{x_{k+1},\rho}}{a}{\cfg{x_{k+1}',\rho'}}}
{\astep
 {\cfg{\csi{\seq{x_1} \conc \ldots \conc
                                   \seq{x_{k+1}} \conc \alpha},\rho}}
 {a}
 {\cfg{\csi{\alpha \conc \seq{x_{k+1}'}},\rho'}}}
\hfill (k \geq 0)
\\
\Rule
{\sdterm{x_1},\ldots,\sdterm{x_k}, \dterm{x_l},
 \astep{\cfg{x_{k+1},\rho}}{a}{\cfg{x_{k+1}',\rho'}}}
{\astep
  {\cfg{\csi{\seq{x_1} \conc \ldots \conc
                                    \seq{x_{k+1}} \conc \alpha},\rho}}
 {a}
 {\cfg{\csi{\alpha \conc \seq{\Dead} \conc
                                              \seq{x_{k+1}'}},\rho'}}}
\hfill \quad (k \geq l > 0)
\\
\Rule
{\sterm{x_1},\ldots,\sterm{x_k}}
{\sterm{\csi{\seq{x_1} \conc \ldots \conc \seq{x_k}}}}
\qquad
\Rule
{\sdterm{x_1},\ldots,\sdterm{x_k}, \dterm{x_l}}
{\dterm{\csi{\seq{x_1} \conc \ldots \conc \seq{x_k}}}}
\hfill (k \geq l > 0)
\\
\Rule
{\astep{\cfg{x,\rho}}{a}{\cfg{x',\rho'}}}
{\astep{\cfg{\std{x},\rho}}{a}{\cfg{\std{x'},\rho'}}}
\qquad
\Rule
{\sdterm{x}}
{\dterm{\std{x}}}
\end{ruletbl}
\end{table}
In Table~\ref{sos-cyclstrint}, $a$ stands for an arbitrary action from
$\BActTau$.

Bisimulation equivalence is also a congruence with respect to the
strategic interleaving operator for cyclic interleaving and the deadlock
at termination operator.
This follows immediately from the fact that the transition rules for
\TA\ constitute a complete transition system specification in the
relaxed panth format (see e.g.~\cite{Mid01a}).
The axioms given in Tables~\ref{axioms-cyclstrint}
and~\ref{axioms-StopToDead} are sound with respect to bisimulation
equivalence.

We have taken the operator $\csiop$ for a unary operator of which the
operand denotes a sequence of threads.
This matches well with the intuition that an interleaving strategy such
as cyclic interleaving operates on a thread vector.
We can look upon the operator $\csiop$ as if there is actually an
$n$-ary operator, of which the operands denote threads, for every
$n \in \Nat$.
From Section~\ref{sect-projlim-TC}, we will freely look upon the
operator $\csiop$ in this way for the purpose of more concise expression
of definitions and results concerning the projective limit model for the
thread calculus presented in this paper.

\section{Interaction between Threads and Services}
\label{sect-TAtsc}

A thread may make use of services.
That is, a thread may perform certain actions for the purpose of having
itself affected by a service that takes those actions as commands to be
processed.
At completion of the processing of an action, the service returns a
reply value to the thread.
The reply value determines how the thread proceeds.
In this section, we extend TA to a theory about threads,
multi-threading, and this kind interaction between threads and services.

It is assumed that a fixed but arbitrary set of \emph{foci} $\Foci$ and
a fixed but arbitrary set of \emph{methods} $\Meth$ have been given.
For  the set of basic actions $\BAct$, we take the set
$\FM = \set{f.m \where f \in \Foci, m \in \Meth}$.
Each focus plays the role of a name of a service provided by the
execution environment that can be requested to process a command.
Each method plays the role of a command proper.
Performing a basic action $f.m$ is taken as making a request to the
service named $f$ to process the command $m$.

In order to extend \TA\ to a theory about threads, multi-threading, and
the above-mentioned kind of interaction between threads and services, we
introduce, for each $f \in \Foci$, a binary
\emph{thread-service composition} operator $\use{\ph}{f}{\ph}$.
The thread denoted by a closed term of the form $\use{p}{f}{H}$ is the
thread that results from processing all basic actions performed by the
thread denoted by $p$ that are of the form $f.m$ by the service denoted
by $H$.
On processing of a basic action of the form $f.m$, the resulting thread
performs the action $\Tau$ and proceeds on the basis of the reply value
returned to the thread.

A service may be unable to process certain commands.
If the processing of one of those commands is requested by a thread, the
request is rejected and the thread becomes inactive.
In the representation of services, an additional reply value $\Refused$
is used to indicate that a request is rejected.

A service is represented by a function
$\funct{H}{\neseqof{\Meth}}{\set{\True,\False,\Refused}}$ satisfying
$H(\alpha) = \Refused \Implies H(\alpha \conc \seq{m}) = \Refused$ for
all $\alpha \in \neseqof{\Meth}$ and $m \in \Meth$.
This function is called the \emph{reply} function of the service.
We write $\RF$ for the set of all reply functions.
Given a reply function $H \in \RF$ and a method $m \in \Meth$, the
\emph{derived} reply function of $H$ after processing $m$, written
$\derive{m}H$, is defined by
$\derive{m}H(\alpha) = H(\seq{m} \conc \alpha)$.

The connection between a reply function $H$ and the service represented
by it can be understood as follows:
\begin{iteml}
\item
if $H(\seq{m}) \neq \Refused$, the request to process command $m$ is
accepted by the service, the reply is $H(\seq{m})$, and the service
proceeds as $\derive{m}H$;
\item
if $H(\seq{m}) = \Refused$, the request to process command $m$ is
rejected by the service and the service proceeds as a service that
rejects any request.
\end{iteml}
Henceforth, we will identify a reply function with the service
represented by it.

The axioms for the thread-service composition operators are given in
Table~\ref{axioms-use}.%
\begin{table}[!tb]
\caption{Axioms for thread-service composition}
\label{axioms-use}
\begin{eqntbl}
\begin{saxcol}
\use{\Stop}{f}{H} = \Stop                             & & \axiom{TSC1}\\
\use{\Dead}{f}{H} = \Dead                       & & \axiom{TSC2}\\
\use{(\Tau \bapf x)}{f}{H} =
                           \Tau \bapf (\use{x}{f}{H}) & & \axiom{TSC3}\\
\use{(\pcc{x}{g.m}{y})}{f}{H} =
\pcc{(\use{x}{f}{H})}{g.m}{(\use{y}{f}{H})}
                                        & \mif f \neq g & \axiom{TSC4}\\
\use{(\pcc{x}{f.m}{y})}{f}{H} =
\Tau \bapf (\use{x}{f}{\derive{m}H})
                           & \mif H(\seq{m}) = \True    & \axiom{TSC5}\\
\use{(\pcc{x}{f.m}{y})}{f}{H} =
\Tau \bapf (\use{y}{f}{\derive{m}H})
                           & \mif H(\seq{m}) = \False   & \axiom{TSC6}\\
\use{(\pcc{x}{f.m}{y})}{f}{H} = \Dead
                           & \mif H(\seq{m}) = \Refused & \axiom{TSC7}
\end{saxcol}
\end{eqntbl}
\end{table}
In this table, $f$ and $g$ stand for arbitrary foci from $\Foci$ and $m$
stands for an arbitrary method from $\Meth$.
Axioms TSC3 and TSC4 express that the action $\Tau$ and actions of
the form $g.m$, where $f \neq g$, are not processed.
Axioms TSC5 and TSC6 express that a thread is affected by a service
as described above when an action of the form $f.m$ performed by the
thread is processed by the service.
Axiom TSC7 expresses that deadlock takes place when an action to be
processed is not accepted.

Henceforth, we write \TAtsc\ for \TA(\FM) extended with the
thread-service composition operators and the axioms from
Table~\ref{axioms-use}.

\begin{example}
\label{example-TAtsc}
Let $m,m',m'' \in \Meth$, and
let $H$ be a service such that
$H(\alpha \conc \seq{m}) = \True$ if $\#_{m'}(\alpha) > 0$,
$H(\alpha \conc \seq{m}) = \False$ if $\#_{m'}(\alpha) \leq 0$, and
$H(\alpha \conc \seq{m'}) = \True$, for all $\alpha \in \seqof{\Meth}$.
Here $\#_{m'}(\alpha)$ denotes the number of occurrences of $m'$ in
$\alpha$.
Then the following equation is easily derivable from the axioms of
\TAtsc:
\begin{ldispl}
\use
 {(f.m' \bapf
   (\pcc{(f'.m' \bapf \Stop)}{f.m}{(f''.m'' \bapf \Stop)}))}
 {f}{H} =
\Tau \bapf \Tau \bapf f'.m' \bapf \Stop\;.
\end{ldispl}%
This equation shows clearly how the thread denoted by
$f.m' \bapf (\pcc{(f'.m' \bapf \Stop)}{f.m}{(f''.m'' \bapf \Stop)})$
is affected by service $H$:
on the processing of $f.m'$ and $f.m$, these basic actions are turned
into $\Tau$, and the reply value returned by $H$ after the processing of
$f.m$ makes the thread proceed with performing $f'.m'$.
\end{example}

We can prove that each closed term over the signature of \TAtsc\ can be
reduced to a closed term over the signature of $\BTA(\FM)$.
\begin{theorem}[Elimination]
\label{thm-elim-TAtsc}
For all closed terms $p$ over the signature of \TAtsc, there exists a
closed term $q$ over the signature of $\BTA(\FM)$ such that $p = q$ is
derivable from the axioms of \TAtsc.
\end{theorem}

The following are useful properties of the deadlock at termination
operator in the presence of both cyclic interleaving and thread-service
composition which are proved using Theorem~\ref{thm-elim-TAtsc}.
\begin{proposition}
\label{prop-s2d-csi-use}
For all closed terms $p_1,\ldots,p_n$ over the signature of \TAtsc, the
following equations are derivable from the axioms of \TAtsc:
\begin{eqnarray}
\setcounter{equation}{1}
& &
\std{\csi{\seq{p_1} \conc \ldots \conc \seq{p_n}}} =
\csi{\seq{\std{p_1}} \conc \ldots \conc
            \seq{\std{p_n}}}\;,
\label{s2d-1-use}
\\
& &
\std{\std{p_1}} = \std{p_1}\;,
\label{s2d-2-use}
\\
& &
\std{\use{p_1}{f}{H}} = \use{\std{p_1}}{f}{H}\;.
\label{s2d-3-use}
\end{eqnarray}
\end{proposition}

The structural operational semantics of \TAtsc\ is described by the
transition rules given in Tables~\ref{sos-BTA},
\ref{sos-cyclstrint} and~\ref{sos-use}.
\begin{table}[!tb]
\caption{Transition rules for thread-service composition}
\label{sos-use}
\begin{ruletbl}
\Rule
{\astep{\cfg{x,\rho}}{\Tau}{\cfg{x',\rho'}}}
{\astep{\cfg{\use{x}{f}{H},\rho}}{\Tau}{\cfg{\use{x'}{f}{H},\rho'}}}
\qquad
\RuleC
{\astep{\cfg{x,\rho}}{g.m}{\cfg{x',\rho'}}}
{\astep{\cfg{\use{x}{f}{H},\rho}}{g.m}{\cfg{\use{x'}{f}{H},\rho'}}}
{f \neq g}
\\
\RuleC
{\astep{\cfg{x,\rho}}{f.m}{\cfg{x',\rho'}}}
{\astep{\cfg{\use{x}{f}{H},\rho}}{\Tau}
       {\cfg{\use{x'}{f}{\derive{m}H},\rho'}}}
{H(\seq{m}) \neq \Refused,\,
 \tup{f.m,H(\seq{m})} \in \rho(\emptyseq)}
\\
\RuleC
{\astep{\cfg{x,\rho}}{f.m}{\cfg{x',\rho'}}}
{\dterm{\use{x}{f}{H}}}
{H(\seq{m}) = \Refused}
\qquad
\Rule
{\sterm{x}}
{\sterm{\use{x}{f}{H}}}
\qquad
\Rule
{\dterm{x}}
{\dterm{\use{x}{f}{H}}}
\end{ruletbl}
\end{table}
In Table~\ref{sos-use}, $f$ and $g$ stand for arbitrary foci from
$\Foci$ and $m$ stands for an arbitrary method from $\Meth$.

Bisimulation equivalence is also a congruence with respect to the
thread-service composition operators.
This follows immediately from the fact that the transition rules for
these operators are in the path format.
The axioms given in Table~\ref{axioms-use} are sound with respect to
bisimulation equivalence.

Leaving out of consideration that the use operators introduced
in~\cite{BM04c} support special actions for testing whether commands
will be accepted by services, those operators are the same
as the thread-service composition operators introduced in this section.

We end this section with a precise statement of what we mean by a
regular service.
Let $H \in \RF$.
Then the set $\rDelta(H) \subseteq \RF$ is inductively defined by the
following rules:
\begin{iteml}
\item
$H \in \rDelta(H)$;
\item
if $m \in \Meth$ and $H' \in \rDelta(H)$, then
$\derive{m} H' \in \rDelta(H)$.
\end{iteml}
We say that $H$ is a \emph{regular} service if $\rDelta(H)$ is a finite
set.

In Section~\ref{sect-REC}, we need the notion of a regular service in
Proposition~\ref{prop-use-rec}.
In the state-based approach to describe services that will be introduced
in Section~\ref{sect-service-descr}, a service can be described using a
finite set of states if and only if it is regular.

\section{Recursion}
\label{sect-REC}

We proceed to recursion in the current setting.
In this section, $T$ stands for either \BTA, \TA, \TAtsc\ or \TC\
(\TC\ will be introduced in Section~\ref{sect-TC}).
We extend $T$ with recursion by adding variable binding operators and
axioms concerning these additional operators.
We will write $T+\REC$ for the resulting theory.

For each variable $x$, we add a variable binding \emph{recursion}
operator $\fixop{x}$ to the operators of $T$.

Let $t$ be a term over the signature of $T+\REC$.
Then an occurrence of a variable $x$ in $t$ is \emph{free} if the
occurrence is not contained in a subterm of the form $\fix{x}{t'}$.
A variable $x$ is \emph{guarded} in $t$ if each free occurrence of $x$
in $t$ is contained in a subterm of the form $\pcc{t'}{a}{t''}$.

Let $t$ be a term over the signature of $T+\REC$ such that
$\fix{x}{t}$ is a closed term.
Then $\fix{x}{t}$ stands for a solution of the equation $x = t$.
We are only interested in models of $T+\REC$ in which $x = t$ has a
unique solution if $x$ is guarded in $t$.
If $x$ is unguarded in $t$, then $\Dead$ is always one of the
solutions of $x = t$.
We stipulate that $\fix{x}{t}$ stands for $\Dead$ if $x$ is unguarded
in $t$.

We add the axioms for recursion given in Table~\ref{axioms-rec} to the
axioms of $T$.%
\begin{table}[!tb]
\caption{Axioms for recursion}
\label{axioms-rec}
\begin{eqntbl}
\begin{saxcol}
\fix{x}{t} = t \subst{\fix{x}{t}}{x}                 & & \axiom{REC1} \\
y = t \subst{y}{x} \Implies y = \fix{x}{t}
        & \mif x\; \mathrm{guarded}\;  \mathrm{in}\; t & \axiom{REC2} \\
\fix{x}{x} = \Dead                                   & & \axiom{REC3}
\end{saxcol}
\end{eqntbl}
\end{table}
In this table, $t$ stands for an arbitrary term over the signature of
$T+\REC$.
The side-condition added to REC2 restricts the terms for which $t$
stands to the terms in which $x$ is guarded.
For a fixed $t$ such that $\fix{x}{t}$ is a closed term, REC1 expresses
that $\fix{x}{t}$ is a solution of $x = t$ and REC2 expresses that this
solution is the only one if $x$ is guarded in $t$.
REC3 expresses that $\fix{x}{x}$ is the non-unique solution $\Dead$
of the equation $x = x$.

\begin{example}
\label{example-REC}
Let $m,m' \in \Meth$, and
let $H$ be a service such that
$H(\alpha \conc \seq{m}) = \True$ if $\#_{m}(\alpha) > 3$, and
$H(\alpha \conc \seq{m}) = \False$ if $\#_{m}(\alpha) \leq 3$.
Here $\#_{m}(\alpha)$ denotes the number of occurrences of $m$ in
$\alpha$.
Then the following equation is easily derivable from the axioms of
\TAtsc+\REC:
\begin{ldispl}
\use{\fix{x}{\pcc{(f'.m' \bapf \Stop)}{f.m}{x}}}{f}{H} =
\Tau \bapf \Tau \bapf \Tau \bapf \Tau \bapf f'.m' \bapf \Stop\;.
\end{ldispl}%
This equation shows clearly that the thread denoted by
$\fix{x}{\pcc{(f'.m' \bapf \Stop)}{f.m}{x}}$ performs $f.m$ repeatedly
until the reply from service $H$ is $\True$.
\end{example}

Let $t$ and $t'$ be terms over the signature of $T+\REC$ such that
$\fix{x}{t}$ and $\fix{x}{t'}$ are closed terms and $t = t'$ is
derivable by either applying an axiom of $T$ in either direction or
axiom REC1 from left to right.
Then it is straightforwardly proved, using the necessary and sufficient
condition for preservation of solutions given in~\cite{PU01a}, that
$x = t$ and $x = t'$ have the same set of solutions in any model of $T$.
Hence, if $x = t$ has a unique solution, then $x = t'$ has a unique
solution and those solutions are the same.
This justifies a weakening of the side-condition of axiom REC2 in the
case where $\fix{x}{t}$ is a closed term.
In that case, it can be replaced by ``$x$ is guarded in some term $t'$
for which $t = t'$ is derivable by applying axioms of $T$ in either
direction and/or axiom REC1 from left to right''.

Theorem~\ref{thm-elim-TA} states that the strategic interleaving
operator for cyclic interleaving and the deadlock at termination
operator can be eliminated from closed terms over the signature of \TA.
Theorem~\ref{thm-elim-TAtsc} states that beside that the thread-service
composition operators can be eliminated from closed terms over the
signature of \TAtsc.
These theorems do not state anything concerning closed terms over the
signature of \TA+\REC\ or closed terms over the signature of
\TAtsc+\REC.
The following three propositions concern the case where the operand of
the strategic interleaving operator for cyclic interleaving is a
sequence of closed terms over the signature of \BTA+\REC\ of the form
$\fix{x}{t}$, the case where the operand of the deadlock at termination
operator is such a closed term, and the case where the first operand of
a thread-service composition operator is such a closed term.
\begin{proposition}
\label{prop-csi-rec}
Let $t$ and $t'$ be terms over the signature of \BTA\textup{+}\REC\ such
that $\fix{x}{t}$ and $\fix{y}{t'}$ are closed terms.
Then there exists a term $t''$ over the signature of \BTA\textup{+}\REC\
such that
$\csi{\seq{\fix{x}{t}} \conc \seq{\fix{y}{t'}}} = \fix{z}{t''}$
is derivable from the axioms of \TA\textup{+}\REC.
\end{proposition}
\begin{proposition}
\label{prop-s2d-rec}
\sloppy
Let $t$ be a term over the signature of \BTA\textup{+}\REC\ such that
$\fix{x}{t}$ is a closed term.
Then there exists a term $t'$ over the signature of \BTA\textup{+}\REC\
such that $\std{\fix{x}{t}} = \fix{y}{t'}$ is derivable from the
axioms of \TA\textup{+}\REC.
\end{proposition}
\begin{proposition}
\label{prop-use-rec}
Let $t$ be a term over the signature of \BTA\textup{+}\REC\ such that
$\fix{x}{t}$ is a closed term.
Moreover, let $f \in \Foci$ and let $H \in \RF$ be a regular service.
Then there exists a term $t'$ over the signature of \BTA\textup{+}\REC\
such that $\use{\fix{x}{t}}{f}{H} = \fix{y}{t'}$ is derivable from the
axioms of \TAtsc\textup{+}\REC.
\end{proposition}
Propositions~\ref{prop-csi-rec}, \ref{prop-s2d-rec}
and~\ref{prop-use-rec} state that the strategic interleaving operator
for cyclic interleaving, the deadlock at termination operator and the
thread-service composition operators can be eliminated from closed terms
of the form $\csi{\seq{\fix{x}{t}} \conc \seq{\fix{y}{t'}}}$,
$\std{\fix{x}{t}}$ and $\use{\fix{x}{t}}{f}{H}$, where $t$ and
$t'$ are terms over the signature of \BTA+\REC\ and $H$ is a regular
service.
Moreover, they state that the resulting term is a closed term of the
form $\fix{z}{t''}$, where $t''$ is a term over the signature of
\BTA+\REC.
Proposition~\ref{prop-csi-rec} generalizes to the case where the operand
is a sequence of length greater than $2$.

The transition rules for recursion are given in Table~\ref{sos-rec}.%
\begin{table}[!tb]
\caption{Transition rules for recursion}
\label{sos-rec}
\begin{ruletbl}
\Rule
{\astep{\cfg{t \subst{\fix{x}{t}}{x},\rho}}{a}{\cfg{x',\rho'}}}
{\astep{\cfg{\fix{x}{t},\rho}}{a}{\cfg{x',\rho'}}}
\qquad
\Rule
{\sterm{t \subst{\fix{x}{t}}{x}}}
{\sterm{\fix{x}{t}}}
\qquad
\Rule
{\dterm{t \subst{\fix{x}{t}}{x}}}
{\dterm{\fix{x}{t}}}
\qquad
\Rule
{\phantom{\dterm{\fix{x}{x}}}}
{\dterm{\fix{x}{x}}}
\end{ruletbl}
\end{table}
In this table, $x$ and $t$ stand for an arbitrary variable and an
arbitrary term over the signature of $T+\REC$, respectively, such that
$\fix{x}{t}$ is a closed term.
In this table, $a$ stands for an arbitrary action from $\BActTau$.

The transition rules for recursion given in Table~\ref{sos-rec} are not
in the path format.
They can be put in the generalized panth format from~\cite{Mid01a},
which guarantees that bisimulation equivalence is a congruence with
respect to the recursion operators, but that requires generalizations of
many notions that are material to structural operational semantics.
The axioms given in Table~\ref{axioms-rec} are sound with respect to
bisimulation equivalence.

This is the first time that recursion is incorporated in thread algebra
by adding recursion operators.
Usually, it is incorporated by adding constants for solutions of systems
of recursion equations (see e.g.~\cite{BM06a}).
However, that way of incorporating recursion does not go with the
restriction operators that will be introduced in Section~\ref{sect-TC}.

\section{State-Based Description of Services}
\label{sect-service-descr}

In this section, we introduce the state-based approach to describe a
family of services which will be used in Section~\ref{sect-MDS}.
This approach is similar to the approach to describe state machines
introduced in~\cite{BP02a}.

In this approach, a family of services is described by
\begin{itemize}
\item
a set of states $S$;
\item
an effect function $\funct{\eff}{\Meth \x S}{S}$;
\item
a yield function
$\funct{\yld}{\Meth \x S}{\set{\True,\False,\Refused}}$;
\end{itemize}
satisfying the following condition:
\begin{ldispl}
\Exists{s \in S}
 {\Forall{m \in \Meth}
   {{} \\ \quad
    (\yld(m,s) = \Refused \And
     \Forall{s' \in S}
      {(\yld(m,s') = \Refused \Implies \eff(m,s') = s)})}}\;.
\end{ldispl}
The set $S$ contains the states in which the service may be, and the
functions $\eff$ and $\yld$ give, for each method $m$ and state $s$, the
state and reply, respectively, that result from processing $m$ in state
$s$.
By the condition imposed on $S$, $\eff$ and $\yld$, after a request has
been rejected by the service, it gets into a state in which any request
will be rejected.

We define, for each $s \in S$, a cumulative effect function
$\funct{\ceff_s}{\seqof{\Meth}}{S}$ in terms of $s$ and $\eff$ as follows:
\begin{ldispl}
\ceff_s(\emptyseq) = s\;,
\\
\ceff_s(\alpha \conc \seq{m}) = \eff(m,\ceff_s(\alpha))\;.
\end{ldispl}
We define, for each $s \in S$, a service
$\funct{H_s}{\neseqof{\Meth}}{\set{\True,\False,\Refused}}$
in terms of $\ceff_s$ and $\yld$ as follows:
\begin{ldispl}
H_s(\alpha \conc \seq{m}) = \yld(m,\ceff_s(\alpha))\;.
\end{ldispl}
$H_s$ is called the service with \emph{initial state} $s$ described by
$S$, $\eff$ and $\yld$.
We say that $\set{H_s \where s \in S}$ is the \emph{family of services}
described by $S$, $\eff$ and $\yld$.

For each $s \in S$, $H_s$ is a service indeed: the condition imposed on
$S$, $\eff$ and $\yld$ implies that $H_s(\alpha) = \Refused \Implies
H_s(\alpha \conc \seq{m}) = \Refused$ for all
$\alpha \in \neseqof{\Meth}$ and $m \in \Meth$.
It is worth mentioning that $H_s(\seq{m}) = \yld(m,s)$ and
$\derive{m} H_s = H_{\eff(m,s)}$.

\section{Services for Molecular Dynamics}
\label{sect-MDS}

In this section, we describe a family of services which concerns
molecular dynamics.
The formal description given here elaborates on an informal description
of molecular dynamics given in~\cite{BB02a}.

The states of molecular dynamics services resemble collections of
molecules composed of atoms and the methods of molecular dynamics
services transform the structure of molecules like in chemical
reactions.
An atom can have \emph{fields} and each of those fields can contain an
atom.
An atom together with the ones it has links to via fields can be viewed
as a submolecule, and a submolecule that is not contained in a larger
submolecule can be viewed as a molecule.
Thus, the collection of molecules that make up a state can be viewed as
a fluid.
By means of methods, new atoms can be created, fields can be added to
and removed from atoms, and the contents of fields of atoms can be
examined and modified.
A few methods use a \emph{spot} to put an atom in or to get an atom
from.
By means of methods, the contents of spots can be compared and modified
as well.
Creating an atom is thought of as turning an element of a given set of
\emph{proto-atoms} into an atom.
If there are no proto-atoms left, then atoms can no longer be created.

It is assumed that a set $\Spot$ of \emph{spots} and a set $\Field$ of
\emph{fields} have been given.
It is also assumed that a countable set $\PAtom$ of \emph{proto-atoms}
such that $\bot \not\in \PAtom$ and a bijection
$\funct{\proatom}{[1,\card(\PAtom)]}{\PAtom}$ have been given.
Although the set of proto-atoms may be infinite, there exists at any
time only a finite number of atoms.
Each of those atoms has only a finite number of fields.
Modular dynamics services have the following methods:
\begin{itemize}
\item
for each $s \in \Spot$, a \emph{create atom method} $\creatom{s}$;
\item
for each $s,s' \in \Spot$, a \emph{set spot method} $\setspot{s}{s'}$;
\item
for each $s, \in \Spot$, a \emph{clear spot method} $\clrspot{s}$;
\item
for each $s,s' \in \Spot$,
an \emph{equality test method} $\equaltst{s}{s'}$;
\item
for each $s \in \Spot$,
an \emph{undefinedness test method} $\undeftst{s}$;
\item
for each $s \in \Spot$ and $v \in \Field$,
a \emph{add field method} $\addfield{s}{v}$;
\item
for each $s \in \Spot$ and $v \in \Field$,
a \emph{remove field method} $\rmvfield{s}{v}$;
\item
for each $s \in \Spot$ and $v \in \Field$,
a \emph{has field method} $\hasfield{s}{v}$;
\item
for each $s,s' \in \Spot$ and $v \in \Field$,
a \emph{set field method} $\setfield{s}{v}{s'}$;
\item
for each $s,s' \in \Spot$ and $v \in \Field$,
a \emph{get field method} $\getfield{s}{s'}{v}$.
\end{itemize}
We write $\Meth_\md$ for the set of all methods of modular dynamics
services.
It is assumed that $\Meth_\md \subseteq \Meth$.

The states of modular dynamics services comprise the contents of all
spots, the fields of the existing atoms, and the contents of those
fields.
The methods of modular dynamics services can be explained as follows:
\begin{itemize}
\item
$\creatom{s}$:
if an atom can be created, then the contents of spot $s$ becomes a newly
created atom and the reply is $\True$; otherwise, nothing changes and
the reply is $\False$;
\item
$\setspot{s}{s'}$:
the contents of spot $s'$ becomes the same as the contents of spot $s$
and the reply is $\True$;
\item
$\clrspot{s}$:
the contents of spot $s$ becomes undefined and the reply is $\True$;
\item
$\equaltst{s}{s'}$:
if the contents of spot $s$ equals the contents of spot $s'$, then
nothing changes and the reply is $\True$; otherwise, nothing changes and
the reply is $\False$;
\item
$\undeftst{s}$:
if the contents of spot $s$ is undefined, then nothing changes and the
reply is $\True$; otherwise, nothing changes and the reply is $\False$;
\item
$\addfield{s}{v}$:
if the contents of spot $s$ is an atom and $v$ is not yet a field of
that atom, then $v$ is added (with undefined contents) to the fields of
that atom and the reply is $\True$; otherwise, nothing changes and the
reply is $\False$;
\item
$\rmvfield{s}{v}$:
if the contents of spot $s$ is an atom and $v$ is a field of that atom,
then $v$ is removed from the fields of that atom and the reply is
$\True$; otherwise, nothing changes and the reply is $\False$;
\item
$\hasfield{s}{v}$:
if the contents of spot $s$ is an atom and $v$ is a field of that atom,
then nothing changes and the reply is $\True$; otherwise, nothing
changes and the reply is $\False$;
\item
$\setfield{s}{v}{s'}$:
if the contents of spot $s$ is an atom and $v$ is a field of that atom,
then the contents of spot $s'$ becomes the same as the contents of that
field and the reply is $\True$; otherwise, nothing changes and the reply
is $\False$;
\item
$\getfield{s}{s'}{v}$:
if the contents of spot $s'$ is an atom and $v$ is a field of that atom,
then the contents of that field becomes the same as the contents of spot
$s$ and the reply is $\True$; otherwise, nothing changes and the reply
is $\False$.
\end{itemize}
In the explanation given above, wherever we say that the contents of a
spot or field becomes the same as the contents of another spot or field,
this is meant to imply that the former contents becomes undefined if the
latter contents is undefined.

The state-based description of the family of modular dynamics services
is as follows:
\begin{ldispl}
S =
\set{\tup{\sigma,\alpha} \in \nm{SS} \x \nm{AS} \where
     \rng(\sigma) \subseteq \dom(\alpha) \union \set{\bot} \And {}
\quad\quad\quad\;\;
\\ \hfill
         \Forall{a \in \dom(\alpha)}
          {\rng(\alpha(a)) \subseteq
           \dom(\alpha) \union \set{\bot}}} \union \set{\undef}\;,
\end{ldispl}
where
\begin{ldispl}
\begin{aeqns}
\nm{SS} & = & \Spot \to (\PAtom \union \set{\bot})\;,
\eqnsep
\nm{AS} & = &
\Union{A \in \fsetof{(\PAtom)}}
 (A \to
  \Union{F \in \fsetof{(\Field)}} (F \to (\PAtom \union \set{\bot})))\;,
\end{aeqns}
\end{ldispl}
and $\undef \not\in \nm{SS} \x \nm{AS}$;
$s_0$ is some $\tup{\sigma,\alpha} \in S$; and
$\eff$ and $\yld$ are defined in Tables~\ref{eff-mds}
and~\ref{yld-mds}.%
\begin{table}[!tb]
\caption{Effect function for molecular dynamics services}
\label{eff-mds}
\begin{eqntbl}
\begin{axcol}
\eff(\creatom{s},\tup{\sigma,\alpha}) = {} \\ \;\;
\tup{\sigma \owr \maplet{s}{\newatom(\dom(\alpha))},
     \alpha \owr \maplet{\newatom(\dom(\alpha))}{\emptymap}}
 & \mif \newatom(\dom(\alpha)) \neq \bot
\\
\eff(\creatom{s},\tup{\sigma,\alpha}) = \tup{\sigma,\alpha}
 & \mif \newatom(\dom(\alpha)) = \bot
\\
\eff(\setspot{s}{s'},\tup{\sigma,\alpha}) =
\tup{\sigma \owr \maplet{s}{\sigma(s')},\alpha}
\\
\eff(\clrspot{s},\tup{\sigma,\alpha}) =
\tup{\sigma \owr \maplet{s}{\bot},\alpha}
\\
\eff(\equaltst{s}{s'},\tup{\sigma,\alpha}) = \tup{\sigma,\alpha}
\\
\eff(\undeftst{s},\tup{\sigma,\alpha}) = \tup{\sigma,\alpha}
\\
\eff(\addfield{s}{v},\tup{\sigma,\alpha}) = {} \\ \;\;
\tup{\sigma,
     \alpha \owr
     \maplet{\sigma(s)}{\alpha(\sigma(s)) \owr \maplet{v}{\bot}}}
 & \mif \sigma(s) \neq \bot \And v \not\in \dom(\alpha(\sigma(s)))
\\
\eff(\addfield{s}{v},\tup{\sigma,\alpha}) = \tup{\sigma,\alpha}
 & \mif \sigma(s) = \bot \Or v \in \dom(\alpha(\sigma(s)))
\\
\eff(\rmvfield{s}{v},\tup{\sigma,\alpha}) =
\tup{\sigma,
     \alpha \owr
     \maplet{\sigma(s)}{\alpha(\sigma(s)) \dsub \set{v}}}
 & \mif \sigma(s) \neq \bot \And v \in \dom(\alpha(\sigma(s)))
\\
\eff(\rmvfield{s}{v},\tup{\sigma,\alpha}) = \tup{\sigma,\alpha}
 & \mif \sigma(s) = \bot \Or v \not\in \dom(\alpha(\sigma(s)))
\\
\eff(\hasfield{s}{v},\tup{\sigma,\alpha}) = \tup{\sigma,\alpha}
\\
\eff(\setfield{s}{v}{s'},\tup{\sigma,\alpha}) = {} \\ \;\;
\tup{\sigma,
     \alpha \owr
     \maplet
      {\sigma(s)}
      {\alpha(\sigma(s)) \owr \maplet{v}{\sigma(s')}}}
 & \mif \sigma(s) \neq \bot \And v \in \dom(\alpha(\sigma(s)))
\\
\eff(\setfield{s}{v}{s'},\tup{\sigma,\alpha}) = \tup{\sigma,\alpha}
 & \mif \sigma(s) = \bot \Or v \not\in \dom(\alpha(\sigma(s)))
\\
\eff(\getfield{s}{s'}{v},\tup{\sigma,\alpha}) =
\tup{\sigma \owr \maplet{s}{\alpha(\sigma(s'))(v)},\alpha}
 & \mif \sigma(s') \neq \bot \And v \in \dom(\alpha(\sigma(s')))
\\
\eff(\getfield{s}{s'}{v},\tup{\sigma,\alpha}) = \tup{\sigma,\alpha}
 & \mif \sigma(s') = \bot \Or v \not\in \dom(\alpha(\sigma(s')))
\\
\eff(m,\tup{\sigma,\alpha}) = \undef
 & \mif m \not\in \Meth_\md
\\
\eff(m,\undef) = \undef
\end{axcol}
\end{eqntbl}
\end{table}
\begin{table}[!tb]
\caption{Yield function for molecular dynamics services}
\label{yld-mds}
\begin{eqntbl}
\begin{axcol}
\yld(\creatom{s},\tup{\sigma,\alpha}) = \True
 & \mif \newatom(\dom(\alpha)) \neq \bot
\\
\yld(\creatom{s},\tup{\sigma,\alpha}) = \False
 & \mif \newatom(\dom(\alpha)) = \bot
\\
\yld(\setspot{s}{s'},\tup{\sigma,\alpha}) = \True
\\
\yld(\clrspot{s},\tup{\sigma,\alpha}) = \True
\\
\yld(\equaltst{s}{s'},\tup{\sigma,\alpha}) = \True
 & \mif \sigma(s) = \sigma(s')
\\
\yld(\equaltst{s}{s'},\tup{\sigma,\alpha}) = \False
 & \mif \sigma(s) \neq \sigma(s')
\\
\yld(\undeftst{s},\tup{\sigma,\alpha}) = \True
 & \mif \sigma(s) = \bot
\\
\yld(\undeftst{s},\tup{\sigma,\alpha}) = \False
 & \mif \sigma(s) \neq \bot
\\
\yld(\addfield{s}{v},\tup{\sigma,\alpha}) = \True
 & \mif \sigma(s) \neq \bot \And v \not\in \dom(\alpha(\sigma(s)))
\\
\yld(\addfield{s}{v},\tup{\sigma,\alpha}) = \False
 & \mif \sigma(s) = \bot \Or v \in \dom(\alpha(\sigma(s)))
\\
\yld(\rmvfield{s}{v},\tup{\sigma,\alpha}) = \True
 & \mif \sigma(s) \neq \bot \And v \in \dom(\alpha(\sigma(s)))
\\
\yld(\rmvfield{s}{v},\tup{\sigma,\alpha}) = \False
 & \mif \sigma(s) = \bot \Or v \not\in \dom(\alpha(\sigma(s)))
\\
\yld(\hasfield{s}{v},\tup{\sigma,\alpha}) = \True
 & \mif \sigma(s) \neq \bot \And v \in \dom(\alpha(\sigma(s)))
\\
\yld(\hasfield{s}{v},\tup{\sigma,\alpha}) = \False
 & \mif \sigma(s) = \bot \Or v \not\in \dom(\alpha(\sigma(s)))
\\
\yld(\setfield{s}{v}{s'},\tup{\sigma,\alpha}) = \True
 & \mif \sigma(s) \neq \bot \And v \in \dom(\alpha(\sigma(s)))
\\
\yld(\setfield{s}{v}{s'},\tup{\sigma,\alpha}) = \False
 & \mif \sigma(s) = \bot \Or v \not\in \dom(\alpha(\sigma(s)))
\\
\yld(\getfield{s}{s'}{v},\tup{\sigma,\alpha}) = \True
 & \mif \sigma(s') \neq \bot \And v \in \dom(\alpha(\sigma(s')))
\\
\yld(\getfield{s}{s'}{v},\tup{\sigma,\alpha}) = \False
 & \mif \sigma(s') = \bot \Or v \not\in \dom(\alpha(\sigma(s')))
\\
\yld(m,\tup{\sigma,\alpha}) = \Refused
 & \mif m \not\in \Meth_\md
\\
\yld(m,\undef) = \Refused
\end{axcol}
\end{eqntbl}
\end{table}
We use the following notation for functions:
$\dom(f)$ for the domain of the function $f$;
$\rng(f)$ for the range of the function $f$;
$\emptymap$ for the empty function;
$\maplet{d}{r}$ for the function $f$ with $\dom(f) = \set{d}$ such that
$f(d) = r$;
$f \owr g$ for the function $h$ with $\dom(h) = \dom(f) \union \dom(g)$
such that for all $d \in \dom(h)$,\, $h(d) = f(d)$ if
$d \not\in \dom(g)$ and $h(d) = g(d)$ otherwise;
and $f \dsub D$ for the function $g$ with $\dom(g) = \dom(f) \diff D$
such that for all $d \in \dom(g)$,\, $g(d) = f(d)$.
The function
$\funct{\newatom}{\fsetof{(\PAtom)}}{(\PAtom \union \set{\bot})}$
is defined by
\begin{ldispl}
\begin{saeqns}
\newatom(A) & = & \proatom(m + 1) & \mif m  <   \card(\PAtom)\;,
\\
\newatom(A) & = & \bot            & \mif m \geq \card(\PAtom)\;,
\end{saeqns}
\end{ldispl}
where $m = \max \set{n \where \proatom(n) \in A}$.

We write $\MDS$ for the family of modular dynamics services described
above.

Let $\tup{\sigma,\alpha} \in S$, let $s \in \Spot$, let
$a \in \dom(\alpha)$, and let $v \in \dom(\alpha(a))$.
Then $\sigma(s)$ is the contents of spot $s$ if $\sigma(s) \neq \bot$,
$v$ is a field of atom $a$, and $\alpha(a)(v)$ is the contents of field
$v$ of atom $a$ if $\alpha(a)(v) \neq \bot$.
The contents of spot $s$ is undefined if $\sigma(s) = \bot$, and the
contents of field $v$ of atom $a$ is undefined if $\alpha(a)(v) = \bot$.
Notice that $\dom(\alpha)$ is taken for the set of all existing atoms.
Therefore, the contents of each spot, i.e.\ each element of
$\rng(\sigma)$, must be in $\dom(\alpha)$ if the contents is defined.
Moreover, for each existing atom $a$, the contents of each of its
fields, i.e.\ each element of $\rng(\alpha(a))$, must be in
$\dom(\alpha)$ if the contents is defined.
The function $\newatom$ turns proto-atoms into atoms.
After all proto-atoms have been turned into atoms, $\newatom$ yields
$\bot$.
This can only happen if the number of proto-atoms is finite.
Molecular dynamics services get into state $\undef$ when refusing a
request to process a command.

The notation for the methods of molecular dynamics services introduced
in this section has a style which makes the notation $f.m$ less suitable
in the case where $m$ is a method of molecular dynamics services.
Therefore, we will henceforth write $f(m)$ instead of $f.m$ if
$m \in \Meth_\md$.

We conclude this section with a simple example of the use of the methods
of molecular dynamics services.
\begin{example}
\label{example-up-dn}
Consider the threads
\begin{ldispl}
P_{n+1} = \md(\creatom{r}) \bapf \md(\setspot{t}{r}) \bapf Q_n
\end{ldispl}
where
\begin{ldispl}
\begin{aeqns}
Q_0 & = & \Stop\;,
\eqnsep
Q_{i+1} & = &
\md(\setspot{s}{t}) \bapf  \md(\creatom{t}) \bapf
\md(\addfield{s}{\nm{up}}) \bapf \md(\addfield{t}{\nm{dn}}) \bapf {}
\\ & &
\md(\setfield{s}{\nm{up}}{t}) \bapf
\md(\setfield{t}{\nm{dn}}{s}) \bapf Q_i\;.
\end{aeqns}
\end{ldispl}
The processing of all basic actions performed by thread $P_4$ by the
molecular dynamics service of which the initial state is the unique
$\tup{\sigma,\alpha} \in S$ such that $\alpha = \emptymap$ yields the
molecule depicted in Figure~\ref{fig-md}.
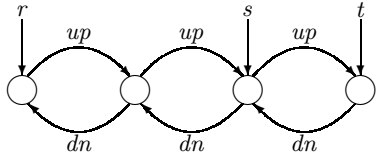
\begin{figure}[!tb]
\begin{center}
\setlength{\unitlength}{0.75mm}
\begin{picture}(80,25)(0,0)
%
\multiput(10,10)(20,0){4}{\circle{5}}
\put(10,22.5){\vector(0,-1){10}}
\put(10,24){\makebox(0,0){$\mathit{r}$}}
\put(50,22.5){\vector(0,-1){10}}
\put(50,24){\makebox(0,0){$\mathit{s}$}}
\put(70,22.5){\vector(0,-1){10}}
\put(70,24.3){\makebox(0,0){$\mathit{t}$}}
\qbezier(11,12.5)(20,22.5)(29,12.5)
\put(27.2,14.5){\vector(1,-1){2}}
\qbezier(31,12.5)(40,22.5)(49,12.5)
\put(47.2,14.5){\vector(1,-1){2}}
\qbezier(51,12.5)(60,22.5)(69,12.5)
\put(67.2,14.5){\vector(1,-1){2}}
\multiput(20,19.5)(20,0){3}{\makebox(0,0){$\mathit{up}$}}
\qbezier(11,7.5)(20,-2.5)(29,7.5)
\put(13.2,5.5){\vector(-1,1){2}}
\qbezier(31,7.5)(40,-2.5)(49,7.5)
\put(33.2,5.5){\vector(-1,1){2}}
\qbezier(51,7.5)(60,-2.5)(69,7.5)
\put(53.2,5.5){\vector(-1,1){2}}
\multiput(20,0.5)(20,0){3}{\makebox(0,0){$\mathit{dn}$}}
\end{picture}
\end{center}
\caption{Molecule yielded by thread $P_4$}
\label{fig-md}
\end{figure}
\end{example}

\section{A Thread Calculus with Molecular Dynamics}
\label{sect-TC}

In this section, \TC\ is introduced.
\TC\ is a version of \TAtsc\ with built-in features of molecular
dynamics and additional operators to restrict the use of certain spots.
Because spots are means of access to atoms, restriction of the use of
certain spots may be needed to prevent interference between threads
in the case where interleaving is involved.

Like in \TAtsc, it is assumed that a fixed but arbitrary set of foci
$\Foci$ and a fixed but arbitrary set of methods $\Meth$ have been
given.
In addition, it is assumed that $\Meth_\md \subseteq \Meth$, spots do
not occur in $m \in \Meth$ if $m \not\in\Meth_\md$, and
$H(\seq{m}) = \Refused$ for all $m \in \Meth_\md$ if  $H \not\in \MDS$.
These additional assumptions express that the methods of molecular
dynamics services are supposed to be built-in and that those methods
cannot be processed by other services.
The last assumption implies that access to atoms is supposed to be
provided by molecular dynamics services only.
Because the operators introduced below to restrict the use of spots
bring along with them the need to rename spots freely, those operators
make it unattractive to have only a limited number of spots available.
Therefore, it is also assumed that $\Spot$ is an infinite set.

Where restriction of their use is concerned, spots are thought of as
names by which atoms are located.
Restriction of the use of spots serves a similar purpose as restriction
of the use of names in the $\pi$-calculus~\cite{MPW92c}.

For each $f \in \Foci$ and $s \in \Spot$, we add a unary
\emph{restriction} operator $\localop{f}{s}$ to the operators of \TAtsc.
The thread denoted by a closed term of the form $\local{f}{s}{p}$ is the
thread denoted by $p$, but the use of spot $s$ is restricted to this
thread as far as basic actions of the form $f.m$ are concerned.
This means that spot $s$ is made a means to access some atom via focus
$f$ that is local to the thread.

The restriction operators of \TC\ are name binding operators of a
special kind.
In $\local{f}{s}{p}$, the occurrence of $s$ in the subscript is a
binding occurrence, but the scope of that occurrence is not simply $p$:
an occurrence of $s$ in $p$ lies within the scope of the binding
occurrence if and only if that occurrence is in a basic action of the
form $f.m$.
As a result, the set of free names of a term, the set of bound names of
a term, and substitutions of names for free occurrences of names in a
term always have a bearing on some focus.
Spot $s$ is a \emph{free name} of term $p$ with respect to focus $f$ if
there is an occurrence of $s$ in $p$ that is in a basic action of the
form $f.m$ that is not in a subterm of the form $\local{f}{s}{p'}$.
Spot $s$ is a \emph{bound name} of term $p$ with respect to focus $f$ if
there is an occurrence of $s$ in $p$ that is in a basic action of the
form $f.m$ that is in a subterm of the form $\local{f}{s}{p'}$.
The \emph{substitution} of spot $s'$ for free occurrences of spot $s$
with respect to focus $f$ in term $p$ replaces in $p$ all occurrences of
$s$ in basic actions of the form $f.m$ that are not in a subterm of the
form $\local{f}{s}{p'}$ by $s'$.

In Appendix~\ref{app-fn-bn-subst}, $\FN^f(p)$, the set of free names of
term $p$ with respect to focus $f$, $\BN^f(p)$, the set of bound names
of term $p$ with respect to focus $f$, and $p \subst{s'}{s}^f$, the
substitution of name $s'$ for free occurrences of name $s$ with respect
to focus $f$ in term $p$, are defined.
We will write $\N(m)$, where $m \in \Meth$, for the set of all names
occurring in $m$.

Par abus de langage, we will henceforth refer to term $p$ as the scope
of the binding occurrence of $s$ in $\local{f}{s}{p}$.

The axioms for restriction are given in Table~\ref{axioms-restriction}.%
\begin{table}[!tb]
\caption{Axioms for restriction}
\label{axioms-restriction}
\begin{eqntbl}
\begin{saxcol}
\local{f}{s}{t} = \local{f}{s'}{t\subst{s'}{s}^f}
                              & \mif s' \not\in \FN^f(t) & \axiom{R1} \\
\local{f}{s}{\Stop} = \Stop                            & & \axiom{R2} \\
\local{f}{s}{\Dead} = \Dead                      & & \axiom{R3} \\
\local{f}{s}{\Tau \bapf x} = \Tau \bapf \local{f}{s}{x}
                                                       & & \axiom{R4} \\
\local{f}{s}{\pcc{x}{g.m}{y}} =
\pcc{\local{f}{s}{x}}{g.m}{\local{f}{s}{y}}
                                         & \mif f \neq g & \axiom{R5} \\
\local{f}{s}{\pcc{x}{f.m}{y}} =
\pcc{\local{f}{s}{x}}{f.m}{\local{f}{s}{y}}
                                  & \mif s \not\in \N(m) & \axiom{R6} \\
\csi{\seq{\local{f}{s}{x}} \conc \alpha} =
\local{f}{s}{\csi{\seq{x} \conc \alpha}}
                          & \mif s \not\in \FN^f(\alpha) & \axiom{R7} \\
\std{\local{f}{s}{x}} = \local{f}{s}{\std{x}}
                                                       & & \axiom{R8} \\
\use{\local{f}{s}{x}}{g}{H} = \local{f}{s}{\use{x}{g}{H}}
                                         & \mif f \neq g & \axiom{R9} \\
\use{\local{f}{s}{x}}{f}{H} = \use{x}{f}{H}
                & \mif H(\seq{\undeftst{s}}) \neq \False & \axiom{R10}\\

\local{f}{s}{\local{g}{s'}{x}} = \local{g}{s'}{\local{f}{s}{x}}
                                                       & & \axiom{R11}
\end{saxcol}
\end{eqntbl}
\end{table}
In this table, $s$ and $s'$ stand for arbitrary spots from $\Spot$, $f$
and $g$ stand for arbitrary foci from $\Foci$, and $t$ stands for an
arbitrary term over the signature of \TC.
%
The crucial axioms are R1, R7, R9 and R10.
Axiom R1 asserts that alpha-convertible restrictions are equal.
Axiom R7 expresses that, in case the scope of a restricted spot is a
thread in a thread vector, the scope can safely be extended to the
strategic interleaving of that thread vector if the restricted spot is
not freely used by the other threads in the thread vector through the
focus concerned.
Axiom R9 expresses that, in case the scope of a restricted spot is a
thread that is composed with a service and the foci concerned are
different, the scope can safely be extended to the thread-service
composition.
Axiom R10 expresses that, in case the scope of a restricted spot is a
thread that is composed with a service and the foci concerned are equal,
the restriction can be raised if the contents of the restricted spot is
undefined -- indicating that it is not in use by any thread to access
some atom.

Axiom R1, together with the assumption that $\Spot$ is infinite, has
an important consequence: in case axiom R7 or axiom R10 cannot be
applied directly because the condition on the restricted spot is not
satisfied, it can always be applied after application of axiom R1.

Next we give a simple example of the use of restriction.
\begin{example}
\label{example-local}
In the expressions $\pcc{p}{\md(\setfield{s}{v}{s'{.}w})}{q}$
and $\pcc{p}{\md(\setfield{s}{v{.}w}{s'})}{q}$, where $p$ and $q$
are terms over the signature of \TC, a get field method is
combined in different ways with a set field method.
This results in expressions that are not terms over the signature of
\TC.
However, these expressions could be considered abbreviations for the
following terms over the signature of \TC:
\begin{ldispl}
\local{\md}{s''}
 {\md(\getfield{s''}{s'}{w}) \bapf
  (\pcc{p}{\md(\setfield{s}{v}{s''})}{q})}\;,
\eqnsep
\local{\md}{s''}
 {\md(\getfield{s''}{s}{v}) \bapf
  (\pcc{p}{\md(\setfield{s''}{w}{s'})}{q})}\;,
\end{ldispl}
where $s'' \not\in \FN^\md(p) \union \FN^\md(q)$.
The importance of the use of restriction here is that it prevents
interference by means of $s''$ in the case where interleaving is
involved, as illustrated by the following derivable equations:
\begin{ldispl}
\csi{\seq{\md(\getfield{s''}{s'}{w}) \bapf
          (\pcc{p}{\md(\setfield{s}{v}{s''})}{q})} \conc
     \seq{\md(\clrspot{s''}) \bapf \Stop}}
\\ \quad {} =
\md(\getfield{s''}{s'}{w}) \bapf \md(\clrspot{s''}) \bapf
(\pcc{p}{\md(\setfield{s}{v}{s''})}{q})\;,
\end{ldispl}
\begin{ldispl}
\csi{\seq{\local{\md}{s''}
           {\md(\getfield{s''}{s'}{w}) \bapf
            (\pcc{p}{\md(\setfield{s}{v}{s''})}{q})}} \conc
     \seq{\md(\clrspot{s''}) \bapf \Stop}}
\\ \quad {} =
\local{\md}{s'''}
 {\md(\getfield{s'''}{s'}{w}) \bapf \md(\clrspot{s''}) \bapf
  (\pcc{p}{\md(\setfield{s}{v}{s'''})}{q})}\;,
\end{ldispl}
where $s''' \not\in \FN^\md(p) \union \FN^\md(q) \union \set{s''}$.
The first equation shows that there is interference if restriction is
not used, whereas the second equation shows that there is no
interference if restriction is used.
Notice that derivation of the second equation requires that axiom R1 is
applied before axiom R7 is applied.
\end{example}

Not every closed term over the signature of \TC\ can be reduced to
a closed term over the signature of $\BTA(\FM)$, e.g.\ a term of
the form $\local{f}{s}{\pcc{p}{f.m}{q}}$, where $p$ and $q$ are
closed terms over the signature of $\BTA(\FM)$, cannot be reduced
further if $s \in \N(m)$.
To elaborate on this remark, we introduce the notion of a basic term.
The set $\cB$ of \emph{basic terms} is inductively defined by the
following rules:
\begin{iteml}
\item
$\Stop,\Dead \in \cB$;
\item
if $p \in \cB$, then $\Tau \bapf p \in \cB$;
\item
if $f \in \Foci$, $m \in \Meth$, and $p,q \in \cB$, then
$\pcc{p}{f.m}{q} \in \cB$;
\item
if $f \in \Foci$, $m \in \Meth$, $s_1,\ldots,s_n \in \N(m)$,
$s_i \neq s_j$ for all $i,j \in [1,n]$ with $i \neq j$, and
$p,q \in \cB$, then
$\local{f}{s_1}{\ldots\local{f}{s_n}{\pcc{p}{f.m}{q}}\ldots} \in \cB$.
\end{iteml}
We can prove that each closed term over the signature of \TC\ can
be reduced to a term from $\cB$.
\begin{theorem}[Elimination]
\label{thm-elim-TC}
For all closed terms $p$ over the signature of \TC, there exists a
term $q \in \cB$ such that $p = q$ is derivable from the axioms of
\TC.
\end{theorem}
\begin{proof}
The proof follows the same line as the proof of
Theorem~\ref{thm-elim-TAtsc} presented in~\cite{BM06a}.
This means that it is a proof by induction on the structure of $p$ in
which some cases boil down to proving a lemma by some form of induction
or another, mostly structural induction again.
Here, we have to consider the additional case
$p \equiv \local{f}{s}{p'}$, where we can restrict ourselves to basic
terms $p'$.
This case is easily proved by structural induction using axioms
R2--R6 and R11.
In the case
$p \equiv \csi{\seq{p'_1} \conc \ldots \conc \seq{p'_k}}$, where
we can restrict ourselves to basic terms $p'_1,\ldots,p'_k$, we have to
consider the additional case
$p'_1 \equiv
 \local{f}{s_1}{\ldots\local{f}{s_n}{\pcc{p''_1}{f.m}{p'''_1}}\ldots}$
with $s_1,\ldots,s_n \in \N(m)$ and $s_i \neq s_j$ for all
$i,j \in [1,n]$ for which $i \neq j$.
After applying axioms R1 and R7 sufficiently many times at the
beginning, this case goes analogous to the case
$p'_1 \equiv \pcc{p''_1}{f.m}{p'''_1}$.
In the case $p \equiv \std{p'}$, where we can restrict ourselves
to basic terms $p'$, we have to consider the additional case
$p' \equiv
 \local{f}{s_1}{\ldots\local{f}{s_n}{\pcc{p''}{f.m}{p'''}}\ldots}$
with $s_1,\ldots,s_n \in \N(m)$ and $s_i \neq s_j$ for all
$i,j \in [1,n]$ for which $i \neq j$.
After applying axiom R8 $n$ times at the beginning, this case goes
analogous to the case $p' \equiv \pcc{p''}{f.m}{p'''}$.
In the case $p \equiv \use{p'}{f}{H}$, where we can restrict ourselves
to basic terms $p'$, we have to consider the additional case
$p' \equiv
 \local{g}{s_1}{\ldots\local{g}{s_n}{\pcc{p''}{g.m}{p'''}}\ldots}$
with $s_1,\ldots,s_n \in \N(m)$ and $s_i \neq s_j$ for all
$i,j \in [1,n]$ for which $i \neq j$.
After applying axiom R9 or axioms R1 and R10 sufficiently many times at
the beginning, this case goes analogous to the case
$p' \equiv \pcc{p''}{g.m}{p'''}$.
\qed
\end{proof}
The following proposition, concerning the cyclic interleaving of a
thread vector of length $1$ in the presence of thread-service
composition and restriction, is easily proved using
Theorem~\ref{thm-elim-TC}.
\begin{proposition}
\label{prop-csi-use-local}
For all closed terms $p$ over the signature of \TC, the equation
$\csi{\seq{p}} = p$ is derivable from the axioms of \TC.
\end{proposition}
\begin{proof}
The proof follows the same line as the proof of
Proposition~\ref{prop-csi} presented in~\cite{BM06a}.
This means that it is a simple proof by induction on the structure of
$p$.
We have to consider the additional case
$p \equiv
 \local{f}{s_1}{\ldots\local{f}{s_n}{\pcc{p'}{f.m}{p''}}\ldots}$
with $s_1,\ldots,s_n \in \N(m)$ and $s_i \neq s_j$ for all
$i,j \in [1,n]$ for which $i \neq j$.
This case goes similar to the case $p \equiv \pcc{p'}{f.m}{p''}$.
Axioms R1 and R7 are applied sufficiently many times at the beginning
and at the end.
\qed
\end{proof}
The following are useful properties of the deadlock at termination
operator in the presence of thread-service composition and restriction
which are proved using Theorem~\ref{thm-elim-TC}.
\begin{proposition}
\label{prop-s2d-csi-use-local}
For all closed terms $p_1,\ldots,p_n$ over the signature of \TC,
the following equations are derivable from the axioms of \TC:
\begin{eqnarray}
\setcounter{equation}{1}
& &
\std{\csi{\seq{p_1} \conc \ldots \conc \seq{p_k}}} =
\csi{\seq{\std{p_1}} \conc \ldots \conc
            \seq{\std{p_k}}}\;,
\label{s2d-1-use-local}
\\
& &
\std{\std{p_1}} = \std{p_1}\;,
\label{s2d-2-use-local}
\\
& &
\std{\use{p_1}{f}{H}} = \use{\std{p_1}}{f}{H}\;.
\label{s2d-3-use-local}
\end{eqnarray}
\end{proposition}
\begin{proof}
The proof follows the same line as the proof of
Proposition~\ref{prop-s2d-csi-use} presented in~\cite{BM06a}.
This means that equation~(\ref{s2d-1-use-local}) is proved by induction
on the sum of the depths plus one of $p_1,\ldots,p_k$ and case
distinction on the structure of $p_1$, and that
equations~(\ref{s2d-2-use-local}) and~(\ref{s2d-3-use-local}) are proved
by induction on the structure of $p_1$.
For each of the equations, we have to consider the additional case
$p_1 \equiv
 \local{f}{s_1}{\ldots\local{f}{s_n}{\pcc{p'_1}{f.m}{p''_1}}\ldots}$
with $s_1,\ldots,s_n \in \N(m)$ and $s_i \neq s_j$ for all
$i,j \in [1,n]$ for which $i \neq j$.
For each of the equations, this case goes similar to the case
$p_1 \equiv \pcc{p'_1}{f.m}{p''_1}$.
In case of equation~(\ref{s2d-1-use-local}), axioms R1 and R7 are
applied sufficiently many times at the beginning and at the end.
In case of equation~(\ref{s2d-2-use-local}), axiom R8 is applied $n$
times at the beginning and at the end.
In case of equation~(\ref{s2d-3-use-local}), axiom R9 or axioms R1 and
R10 are applied sufficiently many times at the beginning and at the end.
\qed
\end{proof}

\begin{proposition}
\label{prop-local-rec}
\sloppy
Let $t$ be a term over the signature of \BTA\textup{+}\REC\ such that
$\fix{x}{t}$ is a closed term.
Then there exists a term $t'$ over the signature of
\BTA\textup{+}\REC\
such that $\local{f}{s}{\fix{x}{t}} = \fix{y}{t'}$ is derivable from
the axioms of \TC\textup{+}\REC\ provided for all actions $g.m$
occurring in $t$ either $f \neq g$ or $s \not\in \N(m)$.
\end{proposition}
\begin{proof}
The proof follows the same line as the proofs of
Propositions~\ref{prop-csi-rec}--\ref{prop-use-rec} presented
in~\cite{BM06a}.
\qed
\end{proof}

We refrain from providing a structural operational semantics of \TC.
In the case where we do not deviate from the style of structural
operational semantics adopted for \BTA, \TA\ and \TAtsc, the obvious way
to deal with restriction involves the introduction of bound actions,
together with a scope opening transition rule (for restriction) and a
scope closing transition rule (for thread-service\linebreak[2]
composition), like in~\cite{MPW92c}.
This would complicate matters to such an extent that the structural
operational semantics of \TC\ would add only marginally to a better
understanding.
In Section~\ref{sect-TCf}, we will adapt the strategic interleaving
operator for cyclic interleaving such that it supports a basic form of
thread forking.
In the presence of thread forking, it is even more complicated to deal
with restriction in a structural operational semantics because the name
binding involved becomes more dynamic.

\section{Projection and the Approximation Induction Principle}
\label{sect-AIP}

Each closed term over the signature of \TC\ denotes a finite thread,
i.e.\ a thread of which the length of the sequences of actions that it
can perform is bounded.
However, not each closed term over the signature of \TC+\REC\ denotes a
finite thread: recursion gives rise to infinite threads.
Closed terms over the signature of \TC+\REC\ that denote the same
infinite thread cannot always be proved equal by means of the axioms of
\TC+\REC.
In this section, we introduce the approximation induction principle to
reason about infinite threads.

The approximation induction principle, \AIP\ in short, is based on the
view that two threads are identical if their approximations up to any
finite depth are identical.
The approximation up to depth $n$ of a thread is obtained by cutting it
off after performing a sequence of actions of length $n$.

\AIP\ is the infinitary conditional equation given in
Table~\ref{axioms-AIP}.
Here, following~\cite{BL02a}, approximation up to depth $n$ is phrased
in terms of a unary \emph{projection} operator $\projop{n}$.
The axioms for the projection operators are given in
Table~\ref{axioms-pin}.%
\begin{table}[!tb]
\caption{Approximation induction principle}
\label{axioms-AIP}
\begin{eqntbl}
\begin{axcol}
\AND_{n \geq 0} \proj{n}{x} = \proj{n}{y} \Implies x = y   & \axiom{AIP}
\end{axcol}
\end{eqntbl}
\end{table}
\begin{table}[!tb]
\caption{Axioms for projection operators}
\label{axioms-pin}
\begin{eqntbl}
\begin{axcol}
\proj{0}{x} = \Dead                                      & \axiom{P0} \\
\proj{n+1}{\Stop} = \Stop                                & \axiom{P1} \\
\proj{n+1}{\Dead} = \Dead                                & \axiom{P2} \\
\proj{n+1}{\pcc{x}{a}{y}} =
                       \pcc{\proj{n}{x}}{a}{\proj{n}{y}} & \axiom{P3} \\
\proj{n+1}{\local{f}{s}{x}} =
                             \local{f}{s}{\proj{n+1}{x}} & \axiom{P4}
\end{axcol}
\end{eqntbl}
\end{table}
In this table, $a$ stands for an arbitrary action from $\BActTau$, $s$
stands for an arbitrary spot from $\Spot$, and $f$ stands for an
arbitrary focus from $\Foci$.

Let $T$ stand for either \TC\ or \TC+\REC.
Then we will write $T$+\PR\ for $T$ extended with the projections
operators $\projop{n}$ and axioms P0--P4, and we will write $T$+\AIP\
for $T$ extended with the projections operators $\projop{n}$, axioms
P0--P4, and axiom \AIP.

\AIP\ holds in the projective limit models for \TC\ and \TC+\REC\ that
will be constructed in Sections~\ref{sect-projlim-TC}
and~\ref{sect-projlim-REC}, respectively.
Axiom REC2 is derivable from the axioms of \TC, axiom REC1 and \AIP.

Not every closed term over the signature of \TC+\REC\ can be reduced to
a basic term.
However, we can prove that, for each closed term $p$ over the signature
of \TC+\REC, for each $n \in \Nat$, $\proj{n}{p}$ can be reduced to a
basic term.

First, we introduce the notion of a first-level basic term.
Let $\cC$ be the set of all closed term over the signature of
\TC+\REC+\PR.
Then the set $\cB^1$ of \emph{first-level basic terms} is inductively
defined by the following rules:
\begin{iteml}
\item
$\Stop,\Dead \in \cB^1$;
\item
if $p \in \cC$, then $\Tau \bapf p \in \cB^1$;
\item
if $f \in \Foci$, $m \in \Meth$, and $p,q \in \cC$, then
$\pcc{p}{f.m}{q} \in \cB^1$;
\item
if $f \in \Foci$, $m \in \Meth$, $s_1,\ldots,s_n \in \N(m)$,
$s_i \neq s_j$ for all $i,j \in [1,n]$ with $i \neq j$, and
$p,q \in \cC$, then
$\local{f}{s_1}{\ldots\local{f}{s_n}{\pcc{p}{f.m}{q}}\ldots} \in \cB^1$.
\end{iteml}
Every closed term over the signature of \TC+\REC+\PR\ can be reduced to
a first-level basic term.
\begin{proposition}
\label{prop-TC+REC-elim}
For all closed terms $p$ over the signature of \textup{\TC+\REC+\PR},
there exists a term $q \in \cB^1$ such that $p = q$ is derivable from
the axioms of \textup{\TC+\REC+\PR}.
\end{proposition}
\begin{proof}
This is easily proved by induction on the structure of $p$, and in the
case $p \equiv \csi{\seq{p'_1} \conc \ldots \conc \seq{p'_k}}$ by
induction on $k$ and case distinction on the structure of $p'_1$.
\qed
\end{proof}
Proposition~\ref{prop-TC+REC-elim} is used in the proof of the following
theorem.
\begin{theorem}
\label{thm-elim-TC+REC}
For all closed terms $p$ over the signature of \textup{\TC+\REC},
for all $n \in \Nat$, there exists a term $q \in \cB$ such that
$\proj{n}{p} = q$ is derivable from the axioms of \textup{\TC+\REC+\PR}.
\end{theorem}
\begin{proof}
By Proposition~\ref{prop-TC+REC-elim}, it is sufficient to prove that,
for all closed terms $p \in \cB^1$, for all $n \in \Nat$, there exists a
term $q \in \cB$ such that $\proj{n}{p} = q$ is derivable from the
axioms of \textup{\TC+\REC+\PR}.
This is easily proved by induction on $n$ and case distinction on the
structure of $p$.
\qed
\end{proof}

\section{Thread Forking}
\label{sect-TCf}

In this section, we adapt the strategic interleaving operator for cyclic
interleaving such that it supports a basic form of thread forking.
We will do so like in~\cite{BM04c}.

We add the ternary \emph{forking postconditional composition} operator
$\pcc{\ph}{\NewThr{\ph}}{\ph}$ to the operators of \TC.
Like action prefixing, we introduce \emph{forking prefixing} as an
abbreviation: $\NewThr{p} \bapf q$, where $p$ and $q$ are terms over the
signature of \TC\ with thread forking, abbreviates
$\pcc{q}{\NewThr{p}}{q}$.
Henceforth, the postconditional composition operators introduced in
Section~\ref{sect-BTA} will be called non-forking postconditional
composition operators.

The forking postconditional composition operator has the same shape as
non-forking postconditional composition operators.
Formally, no action is involved in forking postconditional composition.
However, for an operational intuition, in $\pcc{p}{\NewThr{r}}{q}$,
$\NewThr{r}$ can be considered a thread forking action.
It represents the act of forking off thread $r$.
Like with real actions, a reply is produced.
We consider the case where forking off a thread will never be blocked or
fail.
In that case, it always leads to the reply $\True$.
The action $\Tau$ is left as a trace of forking off a thread.
In~\cite{BM04c}, we treat several interleaving strategies for threads
that support a basic form of thread forking.
Those interleaving strategies deal with cases where forking may be
blocked and/or may fail.
All of them can easily be adapted to the current setting.
In~\cite{BM04c}, $\NewThr{r}$ was formally considered a thread forking
action.
We experienced afterwards that this leads to unnecessary complications
in expressing definitions and results concerning the projective limit
model for the thread algebra developed in this paper
(see Section~\ref{sect-projlim-TC}).

The axioms for \TC\ with thread forking, written \TCf, are the axioms
of \TC\ and axioms CSI6, S2D5, TSC8 and R12 from
Table~\ref{axioms-forking}.%
\begin{table}[!tb]
\caption{Additional axioms for thread forking}
\label{axioms-forking}
\begin{eqntbl}
\begin{axcol}
\csi{\seq{\pcc{x}{\NewThr{z}}{y}}\conc \alpha} =
\Tau \bapf \csi{\alpha \conc \seq{z} \conc \seq{x}}    & \axiom{CSI6} \\
\std{\pcc{x}{\NewThr{z}}{y}} =
\pcc{\std{x}}{\NewThr{\std{z}}}{\std{y}}               & \axiom{S2D5} \\
\use{(\pcc{x}{\NewThr{z}}{y})}{f}{H} =
\pcc{(\use{x}{f}{H})}{\NewThr{\use{z}{f}{H}}}{(\use{y}{f}{H})}
                                                       & \axiom{TSC8} \\
\local{f}{s}{\pcc{x}{\NewThr{z}}{y}} =
\pcc{\local{f}{s}{x}}{\NewThr{\local{f}{s}{z}}}{\local{f}{s}{y}}
                                                       & \axiom{R12} \\
\proj{n+1}{\pcc{x}{\NewThr{z}}{y}} =
  \pcc{\proj{n}{x}}{\NewThr{\proj{n}{z}}}{\proj{n}{y}} & \axiom{P5}
\end{axcol}
\end{eqntbl}
\end{table}
The axioms for \TC+\AIP\ with thread forking, written \TCf+\AIP, are
the axioms of \TC\ and axioms CSI6, S2D5, TSC8, R12 and P5 from
Table~\ref{axioms-forking}.

Recursion is added to \TCf\ as it is added to \BTA, \TA, \TAtsc\ and
\TC\ in Section~\ref{sect-REC}, taking the following adapted definition
of guardedness of variables in terms: a variable $x$ is guarded in a
term $t$ if each free occurrence of $x$ in $t$ is contained in a subterm
of the form $\pcc{t'}{a}{t''}$ or $\pcc{t'}{\NewThr{t'''}}{t''}$.

Not all results concerning the strategic interleaving operator for
cyclic interleaving go through if this basic form of thread forking is
added.
Theorems~\ref{thm-elim-TC} and~\ref{thm-elim-TC+REC} go through if we
add the following rule to the inductive definition of $\cB$ given in
Section~\ref{sect-TC}:
if $p,q,r \in \cB$, then $\pcc{p}{\NewThr{r}}{q} \in \cB$.
Proposition~\ref{prop-TC+REC-elim} goes through if we add the following
rule to the inductive definition of $\cB^1$ given in
Section~\ref{sect-AIP}:
if $p,q,r \in \cC$, then $\pcc{p}{\NewThr{r}}{q} \in \cB^1$.
Proposition~\ref{prop-csi-use-local} and the first part of
Proposition~\ref{prop-s2d-csi-use-local} go through for closed terms in
which the forking postconditional composition operator does not occur
only.
Proposition~\ref{prop-csi-rec} goes through for terms in which the
forking postconditional composition operator does not occur.
It is an open problem whether Proposition~\ref{prop-csi-rec} goes
through for terms in which the forking postconditional composition
operator does occur.

The transition rules for cyclic interleaving with thread forking in the
absence of restriction are given in Tables~\ref{sos-cyclstrint}
and~\ref{sos-forking}.%
\begin{table}[!tb]
\caption{Additional transition rules for thread forking}
\label{sos-forking}
\begin{ruletbl}
\Rule
{\phantom{\rho(\xi)(\xi) = \False}}
{\astep{\cfg{\pcc{x}{\NewThr{p}}{y},\rho}}{\NewThr{p}}{\cfg{x,\rho}}}
\\
\Rule
{\sterm{x_1},\ldots,\sterm{x_k},
 \astep{\cfg{x_{k+1},\rho}}{\NewThr{y}}{\cfg{x_{k+1}',\rho'}}}
{\astep
 {\cfg{\csi{\seq{x_1} \conc \ldots \conc
                                   \seq{x_{k+1}} \conc \alpha},\rho}}
 {\Tau}
 {\cfg{\csi{\alpha \conc \seq{y} \conc \seq{x_{k+1}'}},\rho'}}}
\hfill (k \geq 0)
\\
\Rule
{\sdterm{x_1},\ldots,\sdterm{x_k}, \dterm{x_l},
 \astep{\cfg{x_{k+1},\rho}}{\NewThr{y}}{\cfg{x_{k+1}',\rho'}}}
{\astep
  {\cfg{\csi{\seq{x_1} \conc \ldots \conc
                                    \seq{x_{k+1}} \conc \alpha},\rho}}
 {\Tau}
 {\cfg{\csi{\alpha \conc \seq{\Dead} \conc \seq{y} \conc
                                              \seq{x_{k+1}'}},\rho'}}}
\hfill \quad (k \geq l > 0)
\end{ruletbl}
\end{table}
Here, we use a binary relation
$\smash{\astep{\cfg{\ph,\rho}}{\alpha}{\cfg{\ph,\rho'}}}$ for each
$\alpha \in \BActTau \union
 \set{\NewThr{p} \where
      p \mathrm{\,closed\, term\, over\, signature\, of\,} \mbox{\TCf}}$
and $\rho,\rho' \in \ExEnv$.
Bisimulation equivalence is a congruence with respect to cyclic
interleaving with thread forking.
The transition labels containing terms do not complicate matters
because there are no volatile operators involved
(see e.g.~\cite{MGR05a}).

\section{Modelling a More Advanced Form of Thread Forking}
\label{sect-TCf-Java}

In this section, we use restriction to model a form of thread forking
found in contemporary programming languages such as Java and C\#.
The modelling is divided into two steps.
It is assumed that $\md \in \Foci$, $\this \in \Spot$, and
$\acti \in \Field$.

Firstly, we introduce expressions of the form
$\NewThrIL{s}{s'}{p} \bapf q$, where $p$ and $q$ are terms over the
signature of \TCf+\REC\ such that $s \notin \FN^\md(q)$.

The intuition is that $\NewThrIL{s}{s'}{p} \bapf q$ will not only fork
off $p$, like $\NewThr{p} \bapf q$, but will also have the following
side-effect: a new atom is created which is made accessible by means of
spot $s$ to the thread being forked off and by means of spot $s'$ to the
thread forking off.
The new atom serves as a unique object associated with the thread being
forked off.
The spots $s$ and $s'$ serve as the names available in the thread being
forked off and the thread forking off, respectively, to refer to that
object.
The important issue is that $s$ is meant to be locally available only.

An expression of the form $\NewThrIL{s}{s'}{p} \bapf q$, where $p$ and
$q$ are as above, can be considered an abbreviation for the following
term over the signature of \TCf+\REC:
\begin{ldispl}
\local{\md}{s}
 {\md(\creatom{s}) \bapf \md(\setspot{s'}{s}) \bapf
  \NewThr{p} \bapf q}\;.
\end{ldispl}
Restriction is used here to see to it that $s$ does not become globally
available.

Secondly, we introduce expressions of the form
$\NewThrJava{s}{p} \bapf q$, where $p$ and $q$ are terms over the
signature of \TCf+\REC\ such that $\this \notin \FN^\md(q)$.
The spot $\this$ corresponds with the self-reference \texttt{this} in
Java.

The intuition is that $\NewThrJava{s}{p} \bapf q$ behaves as
$\NewThrIL{\this}{s}{p} \bapf q$, except that it is not till the thread
forking off issues a start command that the thread being forked off
behaves as $p$.
In other words, $\NewThrJava{s}{p} \bapf q$ is closely related to the
form of thread forking that is for instance found in Java, where first a
statement of the form \texttt{AThread s = new AThread} is used to create
a thread object and then a statement of the form \texttt{s.start()} is
used to start the execution of the thread associated with the created
object.

An expression of the form $\NewThrJava{s}{p} \bapf q$, where $p$ and $q$
are as above, can be considered an abbreviation for the following term
over the signature of \TCf+\REC, using the abbreviation introduced
above:
\begin{ldispl}
\NewThrIL{\this}{s}{\fix{x}{\pcc{p}{\md(\hasfield{s}{\acti})}{x}}} \bapf
q\;.
\end{ldispl}
This means that the action $\md(\addfield{s}{\acti})$ can be used in $q$
as start command for $p$, and by that corresponds with the statement
\texttt{s.start()} in Java.

In the remainder of this section, we introduce Java-like thread forking
in a program notation which is close to existing assembly languages and
describe the behaviour produced by programs in this program notation by
means of \TCf+\REC.

A hierarchy of program notations rooted in program algebra is introduced
in~\cite{BL02a}.
One program notation that belongs to this hierarchy is \PGLD, a very
simple program notation which is close to existing assembly languages.
It has absolute jump instructions and no explicit termination
instruction.
Here, we introduce \PGLDf, an extension of \PGLD\ with fork
instructions.

The \emph{primitive instructions} of \PGLDf\ are:
\begin{iteml}
\item
for each $a \in \BAct$, a \emph{basic instruction} $a$;
\item
for each $a \in \BAct$, a \emph{positive test instruction} $\ptst{a}$;
\item
for each $a \in \BAct$, a \emph{negative test instruction} $\ntst{a}$;
\item
for each $l \in \Nat$, an \emph{absolute jump instruction} $\ajmp{l}$;
\item
for each $s \in \Spot$ and $l \in \Nat$,
an \emph{absolute fork instruction} $\afrk{s}{l}$.
\end{iteml}
A \PGLDf\ program has the form $u_1 \concop \ldots \concop u_n$, where
$u_1,\ldots,u_n$ are primitive instructions of \PGLDf.

The intuition is that the execution of a basic action $a$ produces
either $\True$ or $\False$ at its completion.
In the case of a positive test instruction $\ptst{a}$, $a$ is executed
and execution proceeds with the next primitive instruction if $\True$ is
produced.
Otherwise, the next primitive instruction is skipped and execution
proceeds with the primitive instruction following the skipped one.
In the case of a negative test instruction $\ntst{a}$, the role of the
value produced is reversed.
In the case of a basic instruction $a$, execution always proceeds as if
$\True$ is produced.
The effect of an absolute jump instruction $\ajmp{l}$ is that execution
proceeds with the \mbox{$l$-th} instruction of the program concerned.
If $\ajmp{l}$ is itself the $l$-th instruction, deadlock occurs.
At any stage, if there is no instruction to proceed execution with,
termination occurs.

Let $\gA$ be a model of \TCf+\REC+\AIP.
Then the \emph{thread extraction} operation $\extr{\ph}$ gives, for each
\PGLDf\ program $P$, an element from the domain of $\gA$ that represents
the thread produced by $P$.
This operation is defined by
$\extr{u_1 \concop \ldots \concop u_n} =
 \csi{\seq{\sextr{1}{u_1 \concop \ldots \concop u_n}}}$,
where the operation $\sextr{\ph}{\ph}$ is defined by the equations given
in Table~\ref{axioms-thread-extr} (for $u_1,\ldots,u_n$ primitive
instructions of \PGLDf, $i,l \in \Nat$, and $a \in \BAct$)%
\begin{table}[!tb]
\caption{Defining equations for thread extraction}
\label{axioms-thread-extr}
\begin{eqntbl}
\begin{aeqns}
\sextr{i}{u_1 \concop \ldots \concop u_n} & = & \Stop
& \mif \Not 1 \leq i \leq n \\
\sextr{i}{u_1 \concop \ldots \concop u_n} & = &
a \bapf \sextr{i+1}{u_1 \concop \ldots \concop u_n}
& \mif u_i = a \\
\sextr{i}{u_1 \concop \ldots \concop u_n} & = &
\pcc{\sextr{i+1}{u_1 \concop \ldots \concop u_n}}{a}
       {\sextr{i+2}{u_1 \concop \ldots \concop u_n}} \;\;
& \mif u_i = \ptst{a} \\
\sextr{i}{u_1 \concop \ldots \concop u_n} & = &
\pcc{\sextr{i+2}{u_1 \concop \ldots \concop u_n}}{a}
       {\sextr{i+1}{u_1 \concop \ldots \concop u_n}}
& \mif u_i = \ntst{a} \\
\sextr{i}{u_1 \concop \ldots \concop u_n} & = &
\sextr{l}{u_1 \concop \ldots \concop u_n} \;\;
& \mif u_i = \ajmp{l} \\
\sextr{i}{u_1 \concop \ldots \concop u_n} & = &
\NewThrJava{s}{\sextr{l}{u_1 \concop \ldots \concop u_n}} \bapf
\sextr{i+1}{u_1 \concop \ldots \concop u_n} \;\;
& \mif u_i = \afrk{s}{l}
\end{aeqns}
\end{eqntbl}
\end{table}
and the rule that $\sextr{i}{u_1 \concop \ldots \concop u_n} = \Dead$
if $u_i$ is a jump instruction contained in a cyclic chain of jump
instructions.

Two \PGLDf\ programs are considered behavioural equivalent if
$\extr{P} = \extr{Q}$.
We will come back to behavioural equivalence of \PGLDf\ programs in
Section~\ref{sect-PGLDf-eq}.

\section{Projective Limit Model for \TC}
\label{sect-projlim-TC}


In this section, we construct a projective limit model for \TC.
In this model, which covers finite and infinite threads, threads are
represented by infinite sequences of finite approximations.

To express definitions more concisely, the interpretations of the
constants and operators from the signature of \TC\ in the initial model
of \TC\ and the projective limit model of \TC\ are denoted by the
constants and operators themselves.
The ambiguity thus introduced could be obviated by decorating the
symbols, with different decorations for different models, when they are
used to denote their interpretation in a model.
However, in this paper, it is always immediately clear from the context
how the symbols are used.
Moreover, we believe that the decorations are more often than not
distracting.
Therefore, we leave it to the reader to mentally decorate the symbols
wherever appropriate.

The projective limit construction is known as the inverse limit
construction in domain theory, the theory underlying the approach of
denotational semantics for programming languages
(see e.g.~\cite{Sch86a}).
In process algebra, this construction has been applied for the first
time by Bergstra and Klop~\cite{BK84b}.

We will write $\ITC$ for the domain of the initial model of
\TC.
$\ITC$ consists of the equivalence classes of terms from $\cB$ with
respect to the equivalence induced by the axioms of \TC.
In other words, modulo equivalence, $\ITC$ is $\cB$.
Henceforth, we will identify terms from $\cB$ with their equivalence
class where elements of $\ITC$ are concerned.

Each element of $\ITC$ represents a finite thread, i.e.\ a thread of
which the length of the sequences of actions that it can perform is
bounded.
Below, we will construct a model that covers infinite threads as well.
In preparation for that, we define for all $n$ a function that cuts off
finite threads from $\ITC$ after performing a sequence of actions of
length $n$.

For all $n \in \Nat$, we have the \emph{projection} function
$\funct{\projop{n}}{\ITC}{\ITC}$, inductively defined by
\begin{ldispl}
\proj{0}{p} = \Dead\;, \\
\proj{n+1}{\Stop} = \Stop\;, \\
\proj{n+1}{\Dead} = \Dead\;, \\
\proj{n+1}{\pcc{p}{a}{q}} = \pcc{\proj{n}{p}}{a}{\proj{n}{q}}\;, \\
\proj{n+1}{\local{f}{s}{p}} = \local{f}{s}{\proj{n+1}{p}}\;.
\end{ldispl}
For $p \in \ITC$, $\proj{n}{p}$ is called the $n$-th projection of $p$.
It can be thought of as an approximation of $p$.
If $\proj{n}{p} \neq p$, then $\proj{n+1}{p}$ can be thought of as the
closest better approximation of $p$.
If $\proj{n}{p} = p$, then $\proj{n+1}{p} = p$ as well.
For all $n \in \Nat$, we will write $\ITCn{n}$ for
$\set{\proj{n}{p} \where p \in \ITC}$.

The semantic equations given above to define the projection functions
have the same shape as the axioms for the projection operators
introduced in Section~\ref{sect-AIP}.
We will come back to this at the end of Section~\ref{sect-projlim-REC}.

The properties of the projection operations stated in the following two
lemmas will be used frequently in the sequel.
\begin{lemma}
\label{lemma-min-proj}
For all $p \in \ITC$ and $n,m \in \Nat$,
$\proj{n}{\proj{m}{p}} = \proj{\min \set{n,m}}{p}$.
\end{lemma}
\begin{proof}
This is easily proved by induction on the structure of $p$.
\qed
\end{proof}
\begin{lemma}
\label{lemma-distr-proj}
For all $p_1,\ldots,p_m \in \ITC$ and $n,n_1,\ldots,n_m \in \Nat$ with
$n \leq n_1,\ldots,n_m$:
\begin{eqnarray}
\setcounter{equation}{1}
& &
\proj{n}{\csi{\seq{p_1} \conc \ldots \conc \seq{p_m}}} =
\proj{n}{\csi{\seq{\proj{n_1}{p_1}} \conc \ldots \conc
         \seq{\proj{n_m}{p_m}}}}\;,
\label{distr-proj-1}
\\ & &
\proj{n}{\std{p_1}} = \std{\proj{n}{p_1}}\;,
\label{distr-proj-2}
\\ & &
\proj{n}{\use{p_1}{f}{H}} = \use{\proj{n}{p_1}}{f}{H}\;.
\label{distr-proj-3}
\end{eqnarray}
\end{lemma}
\begin{proof}
Equation~\ref{distr-proj-1} is straightforwardly proved by induction on
$n + m$ and case distinction on the structure of $p_1$.
Equations~\ref{distr-proj-2} and~\ref{distr-proj-3} are easily proved by
induction on the structure of $p_1$.
\qed
\end{proof}

In the projective limit model, which covers finite and infinite threads,
threads are represented by \emph{projective sequences}, i.e.\ infinite
sequences $\projseq{p_n}{n}$ of elements of $\ITC$ such that
$p_n \in \ITCn{n}$ and $p_n = \proj{n}{p_{n+1}}$ for all $n \in \Nat$.
In other words, a projective sequence is a sequence of which successive
components are successive projections of the same thread.
The idea is that any infinite thread is fully characterized by the
infinite sequence of all its finite approximations.
We will write $\PTC$ for
$\set{\projseq{p_n}{n} \where
      \AND_{n \in \Nat}
       (p_n \in \ITCn{n} \And p_n = \proj{n}{p_{n+1}})}$.

The \emph{projective limit model} of \TC\ consists of the
following:
\begin{iteml}
\item
the set $\PTC$, the domain of the projective limit model;
\item
an element of $\PTC$ for each constant of \TC;
\item
an operation on $\PTC$ for each operator of \TC;
\end{iteml}
where those elements of $\PTC$ and operations on $\PTC$ are defined as
follows:
\begin{ldispl}
\begin{aeqns}
\Stopp & = & \projseq{\proj{n}{\Stop}}{n}\;,
\\
\Deadp & = & \projseq{\proj{n}{\Dead}}{n}\;,
\\
\pccp{\projseq{p_n}{n}}{a}{\projseq{q_n}{n}} & = &
\projseq{\proj{n}{\pcc{p_n}{a}{q_n}}}{n}\;,
\\
\csip{\seq{\projseq{{p_1}_n}{n}} \conc \ldots \conc
             \seq{\projseq{{p_m}_n}{n}}} & = &
\projseq{\proj{n}{\csi{\seq{{p_1}_n} \conc \ldots \conc
                             \seq{{p_m}_n}}}}{n}\;,
\\
\stdp{\projseq{p_n}{n}} & = &
\projseq{\proj{n}{\std{p_n}}}{n}\;,
\\
\usep{\projseq{p_n}{n}}{f}{H} & = &
\projseq{\proj{n}{\use{p_n}{f}{H}}}{n}\;,
\\
\localp{f}{s}{\projseq{p_n}{n}} & = &
\projseq{\proj{n}{\local{f}{s}{p_n}}}{n}\;.
\end{aeqns}
\end{ldispl}

Using Lemmas~\ref{lemma-min-proj} and~\ref{lemma-distr-proj}, we
easily prove for $\projseq{p_n}{n},\projseq{q_n}{n} \in \PTC$ and
$\projseq{{p_1}_n}{n},\ldots,\projseq{{p_m}_n}{n} \in \PTC$:
\begin{iteml}
\item
$\proj{n}{\proj{n+1}{\pcc{p_{n+1}}{a}{q_{n+1}}}} =
 \proj{n}{\pcc{p_n}{a}{q_n}}$;
\item
$\proj{n}{\proj{n+1}{\csi{\seq{{p_1}_{n+1}} \conc \ldots \conc
                            \seq{{p_m}_{n+1}}}}} =
 \proj{n}{\csi{\seq{{p_1}_n} \conc \ldots \conc \seq{{p_m}_n}}}$;
\item
$\proj{n}{\proj{n+1}{\std{p_{n+1}}}} =
 \proj{n}{\std{p_n}}$;
\item
$\proj{n}{\proj{n+1}{\use{p_{n+1}}{f}{H}}} =
 \proj{n}{\use{p_n}{f}{H}}$;
\item
$\proj{n}{\proj{n+1}{\local{f}{s}{p_{n+1}}}} =
 \proj{n}{\local{f}{s}{p_n}}$.
\end{iteml}
From this and the definition of $\ITCn{n}$, it follows immediately
that the operations defined above are well-defined, i.e.\ they always
yield elements of $\PTC$.

The initial model can be embedded in a natural way in the projective
limit model: each $p \in \ITC$ corresponds to
$\projseq{\proj{n}{p}}{n} \in \PTC$.
We extend projection to an operation on $\PTC$ by defining
$\proj{m}{\projseq{p_n}{n}} = \projseq{p'_n}{n}$, where
$p'_n = p_n$ if $n < m$ and $p'_n = p_m$ if $n \geq m$.
That is, $\proj{m}{\projseq{p_n}{n}}$ is $p_m$ embedded in $\PTC$ as
described above.
Henceforth, we will identify elements of $\ITC$ with their embedding in
$\PTC$ where elements of $\PTC$ are concerned.

It follows immediately from the construction of the projective limit
model of \TC\ that the axioms of \TC\ form a complete axiomatization of
this model for equations between closed terms.

\section{Metric Space Structure for Projective Limit Model}
\label{sect-projlim-metric}

Following~\cite{Kra87a} to some extent, we make $\PTC$ into a metric
space to establish, using Banach's fixed point theorem, that every
guarded operation $\funct{\phi}{\PTC}{\PTC}$ has a unique fixed point.
This is relevant to the expansion of the projective limit model of \TC\
to the projective limit model of \TC+\REC\ in
Section~\ref{sect-projlim-REC}.

An $m$-ary operation $\phi$ on $\PTC$ is a \emph{guarded} operation
if for all $p_1,\ldots,p_m$, $p'_1,\ldots,p'_m \in \PTC$ and
$n \in \Nat$:
\begin{ldispl}
\proj{n}{p_1} = \proj{n}{p'_1} \And \ldots \And
\proj{n}{p_m} = \proj{n}{p'_m}
\\ \quad {}
\Implies
   \proj{n+1}{\phi(p_1,\ldots,p_m)} =
   \proj{n+1}{\phi(p'_1,\ldots,p'_m)}\;.
\end{ldispl}
We say that $\phi$ is an \emph{unguarded} operation if $\phi$ is not a
guarded operation.

The notion of guarded operation, which originates from~\cite{ST91a},
supersedes the notion of guard used in~\cite{Kra87a}.

In the remainder of this section, as well as in
Sections~\ref{sect-projlim-REC} and~\ref{sect-guarded-rec-eqns}, we
assume known the notions of metric space, completion of a metric space,
dense subset in a metric space, continuous function on a metric space,
limit in a metric space and contracting function on a metric space, and
Banach's fixed point theorem.
The definitions of the above-mentioned notions concerning metric spaces
and Banach's fixed point theorem can, for example, be found
in~\cite{Cro89a}.
In this paper, we will consider ultrametric spaces only.
A metric space $\tup{M,d}$ is an \emph{ultrametric space} if for all
$p,p',p'' \in M$, $d(p,p') \leq \max \set{d(p,p''),d(p'',p')}$.

We define a distance function
$\funct{d}{\PTC \x \PTC}{\Real}$ by
\begin{ldispl}
d(p,p') =
2^{- \min \set{n \in \Nat \where \proj{n}{p} \neq \proj{n}{p'}}}
\quad \mif p \neq p'\;, \\
d(p,p') = 0
\hfill \mif p = p'\;.
\end{ldispl}

It is easy to verify that $\tup{\PTC,d}$ is a metric space.
The following theorem summarizes the basic properties of this metric
space.
\begin{theorem}
\label{thm-ultrametric}
\mbox{}
\begin{enuml}
\item
\label{ultrametric}
$\tup{\PTC,d}$ is an ultrametric space;
\item
\label{completion}
$\tup{\PTC,d}$ is the metric completion of the metric space
$\tup{\ITC,d'}$, where $d'$ is the restriction of $d$ to $\ITC$;
\item
\label{dense}
$\ITC$ is dense in $\PTC$;
\item
\label{continuous}
the operations $\funct{\projop{n}}{\PTC}{\ITCn{n}}$ are continuous;
\item
\label{limit}
for all $p \in \PTC$ and $n \in \Nat$, $d(\proj{n}{p},p) < 2^{-n}$,
hence $\lim_{n \to \infty} \proj{n}{p} = p$.
\end{enuml}
\end{theorem}
\begin{proof}
These properties are general properties of metric spaces constructed in
the way pursued here.
Proofs of Properties~\ref{ultrametric}--\ref{dense} can be found
in~\cite{ST91a}.
A proof of Property~\ref{continuous} can be found in~\cite{Dug66a}.
Property~\ref{limit} is proved as follows.
It follows from Lemma~\ref{lemma-min-proj}, by passing to the limit and
using that the projection operations are continuous and $\ITC$ is dense
in $\PTC$, that $\proj{n}{\proj{m}{p}} = \proj{\min \set{n,m}}{p}$ for
$p \in \PTC$ as well.
Hence,
$\min \set{m \in \Nat \where
           \proj{m}{\proj{n}{p}} \neq \proj{m}{p}} > n$,
and consequently $d(\proj{n}{p},p) < 2^{-n}$.
\qed
\end{proof}
The basic properties given above are used in coming proofs.

The properties of the projection operations stated in the following two
lemmas will be used in the proofs of
Theorems~\ref{thm-opns-nonexpansive} and~\ref{thm-unique-fp} given
below.
\begin{lemma}
\label{lemma-min-proj-limit}
For all $p \in \PTC$ and $n,m \in \Nat$,
$\proj{n}{\proj{m}{p}} = \proj{\min \set{n,m}}{p}$.
\end{lemma}
\begin{proof}
As mentioned above in the proof of Theorem~\ref{thm-ultrametric}, this
lemma follows from Lemma~\ref{lemma-min-proj} by passing to the limit
and using that the projection operations are continuous and $\ITC$ is
dense in $\PTC$.
\qed
\end{proof}
\begin{lemma}
\label{lemma-double-proj}
For all $p_1,\ldots,p_m \in \PTC$ and $n \in \Nat$:
\begin{eqnarray}
\setcounter{equation}{1}
& &
\proj{n}{\pcc{p_1}{a}{p_2}} =
\proj{n}{\pcc{\proj{n}{p_1}}{a}{\proj{n}{p_2}}}\;,
\label{double-proj-1}
\\ & &
\proj{n}{\csi{\seq{p_1} \conc \ldots \conc \seq{p_m}}} =
\proj{n}
 {\csi{\seq{\proj{n}{p_1}} \conc \ldots \conc
              \seq{\proj{n}{p_m}}}}\;,
\label{double-proj-2}
\\ & &
\proj{n}{\std{p_1}} =
\proj{n}{\std{\proj{n}{p_1}}}\;,
\label{double-proj-3}
\\ & &
\proj{n}{\use{p_1}{f}{H}} =
\proj{n}{\use{\proj{n}{p_1}}{f}{H}}\;,
\label{double-proj-4}
\\ & &
\proj{n}{\local{f}{s}{p_1}} =
\proj{n}{\local{f}{s}{\proj{n}{p_1}}}\;.
\label{double-proj-5}
\end{eqnarray}
\end{lemma}
\begin{proof}
It is enough to prove Equations~\ref{double-proj-1}--\ref{double-proj-5}
for $p_1,\ldots,p_m \in \ITC$.
The lemma will then follow by passing to the limit and using that
$\projop{n}$ is continuous and $\ITC$ is dense in $\PTC$.
Equations~\ref{double-proj-1} and~\ref{double-proj-5} follow immediately
from Lemma~\ref{lemma-min-proj} and the definition of $\projop{n}$.
Equations~\ref{double-proj-2}--\ref{double-proj-4} follow immediately
from Lemmas~\ref{lemma-min-proj} and~\ref{lemma-distr-proj}.
\qed
\end{proof}

In the terminology of metric topology, the following theorem states that
all operations in the projective limit model of \TC\ are non-expansive.
This implies that they are continuous, with respect to the metric
topology induced by $d$, in all arguments.
\begin{theorem}
\label{thm-opns-nonexpansive}
For all $p_1,\ldots,p_m,p'_1,\ldots,p'_m \in \PTC$:
\begin{eqnarray}
\setcounter{equation}{1}
& &
d(\pcc{p_1}{a}{p_2},\pcc{p'_1}{a}{p'_2}) \leq
\max \set{d(p_1,p'_1),d(p_2,p'_2)}\;,
\label{non-expansive-1}
\\ & &
\begin{array}[c]{@{}l@{}}
d(\csi{\seq{p_1} \conc \ldots \conc \seq{p_m}},
  \csi{\seq{p'_1} \conc \ldots \conc \seq{p'_m}})
\qquad\qquad\;\;
\\ \hfill {} \leq \max \set{d(p_1,p'_1),\ldots,d(p_m,p'_m)}\;,
\end{array}
\label{non-expansive-2}
\\ & &
d(\std{p_1},\std{p'_1}) \leq d(p_1,p'_1)\;,
\label{non-expansive-3}
\\ & &
d(\use{p_1}{f}{H},\use{p'_1}{f}{H}) \leq d(p_1,p'_1)\;,
\label{non-expansive-4}
\\ & &
d(\local{f}{s}{p_1},\local{f}{s}{p'_1}) \leq d(p_1,p'_1)\;.
\label{non-expansive-5}
\end{eqnarray}
\end{theorem}
\begin{proof}
Let
$k_i = \min \set{n \in \Nat \where \proj{n}{p_i} \neq \proj{n}{p_i'}}$
for $i = 1,2$, and let $k = \min \set{k_1,k_2}$.
Then for all $n \in \Nat$, we have $n < k$ iff
$\proj{n}{p_1} = \proj{n}{p_1'}$ and $\proj{n}{p_2} = \proj{n}{p_2'}$.
From this and Lemma~\ref{lemma-double-proj}, it follows immediately that
$\proj{k-1}{\pcc{p_1}{a}{p_2}} = \proj{k-1}{\pcc{p_1'}{a}{p_2'}}$.
Hence,
$k \leq \min \set{n \in \Nat \where
                  \proj{n}{\pcc{p_1}{a}{p_2}} \neq
                  \proj{n}{\pcc{p_1'}{a}{p_2'}}}$,
which completes the proof for the postconditional composition operators.
The proof for the other operators go analogously.
\qed
\end{proof}

The notion of guarded operation is defined without reference to metric
properties.
However, being a guarded operation coincides with having a metric
property that is highly relevant to the issue of unique fixed points: an
operation on $\PTC$ is a guarded operation iff it is contracting.
This is stated in the following lemma.
\begin{lemma}
\label{lemma-guarded-contracting}
An $m$-ary operation $\phi$ on $\PTC$ is a guarded operation iff
for all $p_1,\ldots,p_m,p'_1,\ldots,p'_m \in \PTC$:
\begin{ldispl}
d(\phi(p_1,\ldots,p_m),\phi(p'_1,\ldots,p'_m)) \leq
\frac{1}{2} \cdot \max \set{d(p_1,p'_1),\ldots,d(p_m,p'_m)}\;.
\end{ldispl}
\end{lemma}
\begin{proof}
Let
$k_i = \min \set{n \in \Nat \where \proj{n}{p_i} \neq \proj{n}{p_i'}}$
for $i = 1,\ldots,m$, and let $k = \min \set{k_1,\ldots,k_m}$.
Then for all $n \in \Nat$, $n < k$ iff $\proj{n}{p_1} = \proj{n}{p_1'}$
and \ldots\ and $\proj{n}{p_m} = \proj{n}{p_m'}$.
From this, the definition of a guarded operation and the definition of
$\projop{0}$, it follows immediately that $\phi$ is a guarded operation
iff for all $n < k+1$,
$\proj{n}{\phi(p_1,\ldots,p_m)} = \proj{n}{\phi(p_1',\ldots,p_m')}$.
Hence, $\phi$ is a guarded operation iff
$k+1 \leq \min \set{n \in \Nat \where
                    \proj{n}{\phi(p_1,\ldots,p_m)} \neq
                    \proj{n}{\phi(p_1',\ldots,p_m')}}$,
which completes the proof.
\qed
\end{proof}
We write $\phi^n$, where $\phi$ is a unary operation on $\PTC$, for the
unary operation on $\PTC$ that is defined by induction on $n$ as
follows: $\phi^{0}(p) = p$ and $\phi^{n+1}(p) = \phi(\phi^{n}(p))$.

We have the following important result about guarded operations.
\begin{theorem}
\label{thm-unique-fp}
Let $\funct{\phi}{\PTC}{\PTC}$ be a guarded operation.
Then $\phi$ has a unique fixed point, i.e.\ there exists a unique
$p \in \PTC$ such that $\phi(p) = p$, and
$\projseq{\proj{n}{\phi^n(\Dead)}}{n}$ is the unique fixed point of
$\phi$.
\end{theorem}
\begin{proof}
We have from Theorem~\ref{thm-ultrametric}.\ref{completion}
that $\tup{\PTC,d}$ is a complete metric space and from
Lemma~\ref{lemma-guarded-contracting} that $\phi$ is contracting.
From this, we conclude by Banach's fixed point theorem that $\phi$ has
a unique fixed point.
It is easily proved by induction on $n$, using
Lemma~\ref{lemma-min-proj-limit} and the definition of guarded
operation, that
$\proj{n}{\proj{n+1}{\phi^{n+1}(\Dead)}} = \proj{n}{\phi^n(\Dead)}$.
From this and the definition of $\ITCn{n}$, it follows that
$\projseq{\proj{n}{\phi^n(\Dead)}}{n}$ is an element of $\PTC$.
Moreover, it is easily proved by case distinction between $n = 0$ and
$n > 0$, using this equation, Lemma~\ref{lemma-min-proj-limit} and the
definition of guarded operation, that
$\proj{n}{\phi(\proj{n}{\phi^n(\Dead)})} =
 \proj{n}{\proj{n}{\phi^{n}(\Dead)}}$.
From this, it follows that $\projseq{\proj{n}{\phi^n(\Dead)}}{n}$ is a
fixed point of $\phi$ by passing to the limit and using that $\phi$ is
continuous and $\ITC$ is dense in $\PTC$ (recall that contracting
operations are continuous).
Because $\phi$ has a unique fixed point,
$\projseq{\proj{n}{\phi^n(\Dead)}}{n}$ must be the unique fixed point
of $\phi$.
\qed
\end{proof}

\section{Projective Limit Model for \TC+\REC}
\label{sect-projlim-REC}

The projective limit model for \TC+\REC\ is obtained by expansion of the
projective limit model for \TC\ with a single operation
$\funct{\fixp}{(\PTC \toi \PTC)}{\PTC}$ for all the recursion
operators.%
\footnote
{Given metric spaces $\tup{D,d}$ and $\tup{D',d'}$, we write
 $D \toi D'$ for the set of all non-expansive functions from
 $\tup{D,d}$ to $\tup{D',d'}$.}

The operation $\fixp$ differs from the other operations by taking
functions from $\PTC$ to $\PTC$ as argument.
In agreement with that, for a given assignment in $\PTC$ for variables,
the operand of a recursion operator is interpreted as a function from
$\PTC$ to $\PTC$.
If the recursion operator $\mathsf{fix}_{x}$ is used, then variable $x$
is taken as the variable representing the argument of the function
concerned.
The interpretation of terms over the signature of \TC+\REC\ will be
formally defined in Section~\ref{sect-guarded-rec-eqns}.

The operation $\fixp$ is defined as follows:
\begin{ldispl}
\begin{aeqns}
\fixp(\phi) & = &
\projseq{\proj{n}{\phi^n(\Dead)}}{n}
\quad \mif \phi\; \mathrm{is\; a\; guarded\; operation}{,}
\phantom{\mathrm{n un}}
\\
\fixp(\phi) & = &
\projseq{\proj{n}{\Dead}}{n}
\hfill \mif \phi\; \mathrm{is\; an\; unguarded\; operation}.
\end{aeqns}
\end{ldispl}

From Theorem~\ref{thm-unique-fp}, we know that every guarded operation
$\funct{\phi}{\PTC}{\PTC}$ has only one fixed point and that
$\projseq{\proj{n}{\phi^n(\Dead)}}{n}$ is that fixed point.
The justification for the definition of $\fixp$ for unguarded operations
is twofold:
\begin{iteml}
\item
a function $\phi$ from $\PTC$ to $\PTC$ that is representable by a term
over the signature of \TC+\REC\ is an unguarded operation only if
$\Dead$ is one of the fixed points of $\phi$;
\item
if $\Dead$ is a fixed point of a function $\phi$ from $\PTC$ to
$\PTC$, then
$\projseq{\proj{n}{\Dead}}{n} =
 \projseq{\proj{n}{\phi^n(\Dead)}}{n}$.
\end{iteml}
This implies that, for all function $\phi$ from $\PTC$ to $\PTC$ that
are representable by a term over the signature of \TC+\REC, $\fixp$
yields a fixed point.
Actually, it is the least fixed point with respect to the approximation
relation $\apx$ that is introduced in Appendix~\ref{app-projlim-cpo}.
There may be unguarded operations in $\PTC \toi \PTC$ for which
$\Dead$ is not a fixed point.
However, those operations are not representable by a term over the
signature of \TC+\REC.

It is straightforward to verify that, for every guarded operation
$\funct{\phi}{\PTC}{\PTC}$,
$\projseq{\proj{n}{\phi^n(\Dead)}}{n} =
 \projseq{\proj{n}{\phi^{k(n)}(\Dead)}}{n}$, where
$k(n) =
 \min \set{k \where
           \proj{n}{\phi^k(\Dead)} = \proj{n}{\phi^{k+1}(\Dead)}}$.
The right-hand side of this equation is reminiscent of the definition of
the operation introduced in~\cite{BK82a} for the selection of a fixed
point in a projective limit model for PA, a subtheory of
\ACP~\cite{BK84b} without communication.

We define a distance function
$\funct{\delta}{(\PTC \toi \PTC) \x (\PTC \toi \PTC)}{\Real}$ by
\begin{ldispl}
\delta(\phi,\psi) = \lub \set{d(\phi(p),\psi(p)) \where p \in \PTC}\;.
\end{ldispl}
The distance function $\delta$ is well-defined because for all
$p,p' \in \PTC$, $\delta(p,p') \leq 2^{-1}$.
It is easy to verify that $\tup{\PTC \toi \PTC,\delta}$ is an
ultrametric space.

The following theorem states that $\fixp$ is non-expansive for guarded
operations.
\begin{theorem}
\label{thm-fix-nonexpansive}
For all $\phi,\psi \in \PTC \toi \PTC$ that are guarded operations:
\begin{ldispl}
d(\fixp(\phi),\fixp(\psi)) \leq \delta(\phi,\psi)\;.
\end{ldispl}
\end{theorem}
\begin{proof}
Let $p = \fixp(\phi)$ and $q = \fixp(\psi)$.
Then $\phi(p) = p$, $\psi(q) = q$ and also
$d(\phi(p),\psi(q)) = d(p,q)$.
We have $d(\phi(p),\phi(q)) \leq \frac{1}{2} \cdot d(p,q)$ by
Lemma~\ref{lemma-guarded-contracting} and
$d(\phi(q),\psi(q)) \leq \delta(\phi,\psi)$ by the definition of
$\delta$.
It follows that
$d(\phi(q),\psi(q)) \leq
 \max \set{\frac{1}{2} \cdot d(p,q),\delta(\phi,\psi)}$.
Hence, because $d(\phi(p),\psi(q)) = d(p,q)$, we have
$d(p,q) \leq \delta(\phi,\psi)$.
That is, $d(\fixp(\phi),\fixp(\psi)) \leq \delta(\phi,\psi)$.
\qed
\end{proof}

Projective limit models of \TC+\AIP\ and \TC+\REC+\AIP\ are simply
obtained by expanding the projective limit models of \TC\ and \TC+\REC\
with the projection operations
$\funct{\projop{n}}{\PTC}{\PTC}$ defined at the end of
Section~\ref{sect-projlim-TC}.

\section{Guarded Recursion Equations}
\label{sect-guarded-rec-eqns}

In this section, following~\cite{Kra87a} to some extent, we introduce
the notions of guarded term and guarded recursion equation and show
that every guarded recursion equation has a unique solution in $\PTC$.
This result is to some extent a side result.
Much of the preparation that has to be done to establish it has been
done in Sections~\ref{sect-projlim-metric} and~\ref{sect-projlim-REC}.
Therefore, it seems like a waste to omit this result.

Supplementary, in Appendix~\ref{app-projlim-cpo}, we make $\PTC$ into a
complete partial ordered set and show, using Tarski's fixed point
theorem, that every recursion equation has a least solution in $\PTC$
with respect to the partial order relation concerned.

It is assumed that a fixed but arbitrary set of variables $\cX$ has been
given.

Let $P \subseteq \PTC$ and let $X \subseteq \cX$.
Then we will write $\PT{}{P}$ for the set of all terms over the
signature of \TC+\REC\ with parameters from $P$ and $\PT{X}{P}$ for the
set of all terms from $\PT{}{P}$ in which no other variables than the
ones in $X$ have free occurrences.%
\footnote
{A term with parameters is a term in which elements of the domain of a
 model are used as constants naming themselves.
 For a justification of this mix-up of syntax and semantics in case only
 one model is under consideration, see e.g.~\cite{Hod93a}.}

The interpretation function
$\funct{\Int{\ph}}{\PT{}{P}}{((\cX \to \PTC) \to \PTC)}$
of terms with parameters from $P \subseteq \PTC$ is defined as follows:
\begin{ldispl}
\begin{aeqns}
\Int{x}(\rho) & = & \rho(x)\;,
\\
\Int{p}(\rho) & = & p\;,
\\
\Int{\Stop}(\rho) & = & \Stop\;,
\\
\Int{\Dead}(\rho) & = & \Dead\;,
\\
\Int{\pcc{t_1}{a}{t_2}}(\rho) & = &
\pcc{\Int{t_1}(\rho)}{a}{\Int{t_2}(\rho)}\;,
\\
\Int{\csi{\seq{t_1} \conc \ldots \conc \seq{t_m}}}(\rho) & = &
\csi{\seq{\Int{t_1}(\rho)} \conc \ldots \conc
            \seq{\Int{t_m}(\rho)}}\;,
\\
\Int{\std{t}}(\rho) & = & \std{\Int{t}(\rho)}\;,
\\
\Int{\use{t}{f}{H}}(\rho) & = & \use{\Int{t}(\rho)}{f}{H}\;,
\\
\Int{\local{s}{f}{t}}(\rho) & = & \local{s}{f}{\Int{t}(\rho)}\;,
\\
\Int{\fix{x}{t}}(\rho) & = & \fixp(\phi)\;,
\\
\multicolumn{3}{@{}l@{}}
{\qquad \mathrm{where}\; \funct{\phi}{\PTC}{\PTC}\;
 \mathrm{is\; defined\; by}\;
 \phi(p) = \Int{t}(\rho \owr \maplet{x}{p})\;.}
\end{aeqns}
\end{ldispl}

The property stated in the following lemma will be used in the proof of
Lemma~\ref{lemma-guarded-term} given below.
\begin{lemma}
\label{lemma-subst-value}
Let $P \subseteq \PTC$, let $t \in \PT{}{P}$,
let $x \in \cX$, let $p \in P$, and let $\funct{\rho}{\cX}{\PTC}$.
Then $\Int{t}(\rho \owr \maplet{x}{p}) = \Int{t\subst{p}{x}}(\rho)$.
\end{lemma}
\begin{proof}
This is easily proved by induction on the structure of $t$.
\qed
\end{proof}

Let $x_1,\ldots,x_n \in \cX$, let $X \subseteq \set{x_1,\ldots,x_n}$,
let $P \subseteq \PTC$, and let $t \in \PT{X}{P}$.
Moreover, let $\funct{\rho}{\cX}{\PTC}$.
Then the \emph{interpretation of $t$ with respect to $x_1,\ldots,x_n$},
written $\Int{t}^{x_1,\ldots,x_n}$, is the unique function
$\funct{\phi}{\PTC^n}{\PTC}$ such that
for all $p_1,\ldots,p_n \in \PTC$,
$\phi(p_1,\ldots,p_n) =
 \Int{t}
  (\rho \owr \maplet{x_1}{p_1} \owr \ldots \owr \maplet{x_n}{p_n})$.

The interpretation of $t$ with respect to $x_1,\ldots,x_n$ is
well-defined because it is independent of the choice of $\rho$.

The notion of guarded term defined below is suggested by the fact,
stated in Lemma~\ref{lemma-guarded-contracting} above, that an operation
on $\PTC$ is a guarded operation iff it is contracting.
The only guarded operations, and consequently contracting operations, in
the projective limit model of \TC+\REC\ are the postconditional
composition operations.
Based upon this, we define the notion of guarded term as follows.

Let $P \subseteq \PTC$.
Then the set $\GT{}{P}$ of \emph{guarded} terms with parameters from $P$
is inductively defined as follows:
\begin{iteml}
\item
if $p \in P$, then
$p \in \GT{}{P}$;
\item
$\Stop,\Dead \in \GT{}{P}$;
\item
if $a \in \BAct$ and $t_1,t_2 \in \PT{}{P}$, then
$\pcc{t_1}{a}{t_2} \in \GT{}{P}$;
\item
if $t_1,\ldots,t_m \in \GT{}{P}$, then
$\csi{\seq{t_1} \conc \ldots \conc \seq{t_m}} \in \GT{}{P}$;
\item
if $t \in \GT{}{P}$, then
$\std{t} \in \GT{}{P}$;
\item
if $f \in \Foci$, $H \in \RF$ and $t \in \GT{}{P}$, then
$\use{t}{f}{H} \in \GT{}{P}$;
\item
if $f \in \Foci$, $s \in \Spot$ and $t \in \GT{}{P}$, then
$\local{s}{f}{t} \in \GT{}{P}$;
\item
if $x \in \cX$, $t \in \GT{}{P}$ and $x$ is guarded in $t$, then
$\fix{x}{t} \in \GT{}{P}$.
\end{iteml}

The following lemma states that guarded terms represent operations on
$\PTC$ that are contracting.
\begin{lemma}
\label{lemma-guarded-term}
Let $x_1,\ldots,x_n \in \cX$, let $X \subseteq \set{x_1,\ldots,x_n}$,
let $P \subseteq \PTC$, and let $t \in \PT{X}{P}$.
Then $t \in \GT{}{P}$ only if for all
$p_1,\ldots,p_n,p'_1,\ldots,p'_n \in \PTC$:
\begin{ldispl}
d(\Int{t}^{x_1,\ldots,x_n}(p_1,\ldots,p_n),
  \Int{t}^{x_1,\ldots,x_n}(p'_1,\ldots,p'_n))
\\ \qquad\qquad\qquad\qquad\qquad\qquad\quad {} \leq
  \frac{1}{2} \cdot \max \set{d(p_1,p'_1),\ldots,d(p_n,p'_n)}\;.
\end{ldispl}
\end{lemma}
\begin{proof}
This is easily proved by induction on the structure of $t$ using
Theorems~\ref{thm-opns-nonexpansive} and~\ref{thm-fix-nonexpansive},
Lemmas~\ref{lemma-guarded-contracting} and~\ref{lemma-subst-value}, and
the fact that the postconditional composition operations are guarded
operations.
\qed
\end{proof}

A \emph{recursion equation} is an equation $x = t$, where $x \in \cX$
and $t \in \PT{\set{x}}{P}$ for some $P \subseteq \PTC$.
A recursion equation $x = t$ is a \emph{guarded} recursion equation if
$t \in \GT{}{P}$ for some $P \subseteq \PTC$.
Let $x = t$ be a recursion equation.
Then $p \in \PTC$ is a \emph{solution} of $x = t$ if
$\Int{t}^x(p) = p$.

We have the following important result about guarded recursion
equations.
\begin{theorem}
\label{thm-unique-solution}
Every guarded recursion equation has a unique solution in the projective
limit model for \TC\textup{+}\REC.
\end{theorem}
\begin{proof}
Let $x \in \cX$, let $P \subseteq \PTC$, and
let $t \in \PT{\set{x}}{P}$ be such that $t \in \GT{}{P}$.
We have from Theorem~\ref{thm-ultrametric}.\ref{completion}
that $\tup{\PTC,d}$ is a complete metric space and from
Lemma~\ref{lemma-guarded-term} that $\Int{t}^{x}$ is contracting.
From this, we conclude by Banach's fixed point theorem that
$\Int{t}^{x}$ has a unique fixed point.
Hence, the guarded recursion equation $x = t$ has a unique solution.
\qed
\end{proof}

The projection operations and the distance function as defined in this
paper match well with our intuitive ideas about finite approximations
of threads and closeness of threads, respectively.
The suitability of the definitions given in this paper is supported by
the fact that guarded operations coincide with contracting operations.
However, it is not at all clear whether adaptations of the definitions
are feasible and will lead to different uniqueness results.

\section{Equality in the Projective Limit Model for \TC+\REC+\AIP}
\label{sect-projlim-eq}

In this section, we determine the position in the arithmetical hierarchy
(the Kleene-Mostowski hierarchy) of the equality relation in the
projective limit model for \TC+\REC+\AIP.

We start with a theorem that bears witness to the strength of the axioms
of \TC+\REC+\AIP.
\begin{theorem}
\label{thm-proj-completeness}
For all closed terms $p,q$ over the signature of \textup{\TC+\REC+\PR}
for which $p = q$ holds in the projective limit model of
\textup{\TC+\REC+\AIP},
for all $n \in \Nat$, $\proj{n}{p} = \proj{n}{q}$ is derivable from the
axioms of \textup{\TC+\REC+\PR}.
\end{theorem}
\begin{proof}
Let $n \in \Nat$, and let $p',q' \in \cB^1$ be such that
$\proj{n}{p} = p'$ and $\proj{n}{q} = q'$ are derivable from the axioms
of \TC+\REC+\PR.
Such terms exist by Theorem~\ref{thm-elim-TC+REC}.
By the soundness of the axioms of \TC+\REC+\PR, $\proj{n}{p} = p'$ and
$\proj{n}{q} = q'$ hold in the projective limit model of \TC+\REC+\AIP.
Moreover, because $p = q$ holds in the projective limit model of
\TC+\REC+\AIP, $\proj{n}{p} = \proj{n}{q}$ holds in the projective limit
model of \TC+\REC+\AIP.
Hence, $p' = q'$ holds in the projective limit model of \TC+\REC+\AIP.
Because the axioms of \TC\ form a complete axiomatization of the
restriction of this model to the signature of \TC\ for equations between
closed terms, $p' = q'$ is derivable from the axioms of \TC.
Hence, $\proj{n}{p} = \proj{n}{q}$ is derivable from the axioms of
\TC+\REC+\PR.
\qed
\end{proof}
By Theorem~\ref{thm-elim-TC+REC}, the reduction of terms $\proj{n}{p}$,
where $p$ is a closed term over the signature of \TC+\REC+\PR, to basic
terms is computable.
Moreover, equality of basic terms is syntactic equality modulo axioms
R1 and R11.
Hence, as a corollary of Theorems~\ref{thm-elim-TC+REC}
and~\ref{thm-proj-completeness}, we have the following decidability
result:
\begin{corollary}
\label{corol-decidability}
For closed terms $p,q$ over the signature of \textup{\TC+\REC+\PR} and
$n \in \Nat$, it is decidable, uniformly in $n$, whether
$\proj{n}{p} = \proj{n}{q}$ holds in the projective limit model of
\textup{\TC+\REC+\AIP}.
\end{corollary}

Corollary~\ref{corol-decidability} leads us to the position in the
arithmetical hierarchy of the equality relation in the projective limit
model of \TC+\REC+\AIP.
Recall that a relation is a $\rSigma^0_0$-relation iff it is a
recursive relation, and that a relation is a $\rPi^0_1$-relation iff it
is a co-recursively enumerable relation (see e.g.~\cite{Sho91a,Men97a}).
\begin{theorem}
\label{thm-co-rec-enum}
Let $\cC$ be the set of all closed terms over the signature of
\textup{\TC+\linebreak[2]\REC+\PR}, and
let ${\cong} \subseteq \cC \x \cC$ be the relation defined by
$p \cong q$ iff $p = q$ holds in the projective limit model of
\textup{\TC+\REC+\AIP}.
Then $\cong$ is a $\rPi^0_1$-relation.
\end{theorem}
\begin{proof}
Let $\nm{Pr} \subseteq \Nat \x \cC \x \cC$ be the relation defined by
$\nm{Pr}(n,p,q)$ iff $\proj{n}{p} = \proj{n}{q}$ holds in the projective
limit model of \TC+\REC+\AIP.
By the definition of this model,
$p \cong q \Iff \Forall{n \in \Nat}{\nm{Pr}(n,p,q)}$
for all $p,q \in \cC$.
Moreover, by Corollary~\ref{corol-decidability}, $\nm{Pr}$ is a
$\rSigma^0_0$-relation.
Hence, $\cong$ is a $\rPi^0_1$-relation.
\qed
\end{proof}

\section{Projective Limit Model for \TC\ with Thread Forking}
\label{sect-projlim-TCf}

The construction of the projective limit model for \TCf\ follows the
same line as the construction of the projective limit model for \TC.
In this section, the construction of the projective limit model for
\TCf\ is outlined.

Recall that the basic terms of \TCf\ include closed terms
$\pcc{p}{\NewThr{r}}{q}$, where $p$, $q$ and $r$ are basic terms
(see Section~\ref{sect-TCf}).
The domain $\ITCp$ of the initial model of \TCf\ consists of the
equivalence classes of basic terms of \TCf.

The projection functions $\funct{\projop{n}}{\ITCp}{\ITCp}$ are the
extensions of the projection functions $\funct{\projop{n}}{\ITC}{\ITC}$
inductively defined by the equations given for
$\funct{\projop{n}}{\ITC}{\ITC}$ in Section~\ref{sect-projlim-TC} and
the following equation:
\begin{ldispl}
\proj{n+1}{\pcc{p}{\NewThr{r}}{q}} =
\pcc{\proj{n}{p}}{\NewThr{\proj{n}{r}}}{\proj{n}{q}}\;.
\end{ldispl}
For all $n \in \Nat$, we will write $\ITCnp{n}$ for
$\set{\proj{n}{p} \where p \in \ITCp}$.
Moreover, we will write $\PTCp$ for
$\set{\projseq{p_n}{n} \where
      \AND_{n \in \Nat}
       (p_n \in \ITCnp{n} \And p_n = \proj{n}{p_{n+1}})}$.

Lemmas~\ref{lemma-min-proj} and~\ref{lemma-distr-proj} go through for
$\ITCp$.

The projective limit model of \TCf\ consists of the following:
\begin{iteml}
\item
the set $\PTCp$, the domain of the projective limit model;
\item
an element of $\PTCp$ for each constant of \TCf;
\item
an operation on $\PTCp$ for each operator of \TCf.
\end{iteml}
Those elements of $\PTCp$ and operations on $\PTCp$, with the exception
of the operation associated with the forking postconditional composition
operator, are defined as in the case of the projective limit model for
\TC.
The ternary operation on $\PTCp$ associated with the forking
postconditional composition operator is defined as follows:
\begin{ldispl}
\pcc{\projseq{p_n}{n}}{\NewThr{\projseq{r_n}{n}}}{\projseq{q_n}{n}} =
\projseq{\proj{n}{\pcc{p_n}{\NewThr{r_n}}{q_n}}}{n}\;.
\end{ldispl}
Using Lemma~\ref{lemma-min-proj}, we easily prove that for
$\projseq{p_n}{n},\projseq{q_n}{n},\projseq{r_n}{n} \in \PTCp$:
\begin{ldispl}
\proj{n}{\proj{n+1}{\pcc{p_{n+1}}{\NewThr{r_{n+1}}}{q_{n+1}}}} =
\proj{n}{\pcc{p_n}{\NewThr{r_n}}{q_n}}\;.
\end{ldispl}
From this and the definition of $\ITCnp{n}$, it follows immediately
that the operation defined above always yields elements of $\PTCp$.

Lemma~\ref{lemma-min-proj-limit} goes through for $\PTCp$.
Lemma~\ref{lemma-double-proj} goes through for $\PTCp$ as well; and we
have in addition that for all $p_1,p_2,p_3 \in \PTCp$ and $n \in \Nat$:
\begin{ldispl}
\proj{n}{\pcc{p_1}{\NewThr{p_3}}{p_2}} =
\proj{n}{\pcc{\proj{n}{p_1}}{\NewThr{\proj{n}{p_3}}}{\proj{n}{p_2}}}\;.
\end{ldispl}
\sloppy
Theorem~\ref{thm-opns-nonexpansive} goes through for $\PTCp$; and we
have in addition that for all $p_1,p_2,p_3,p'_1,p'_2,p'_3 \in \PTCp$:
\begin{ldispl}
d(\pcc{p_1}{\NewThr{p_3}}{p_2},\pcc{p'_1}{\NewThr{p'_3}}{p'_2})
 \leq \max \set{d(p_1,p'_1),d(p_2,p'_2),d(p_3,p'_3)}\;.
\end{ldispl}
Lemma~\ref{lemma-guarded-contracting} and Theorem~\ref{thm-unique-fp} go
through for $\PTCp$.

The projective limit model of \TCf+\REC\ is obtained by expansion of the
projective limit model of \TCf\ with a single operation
$\funct{\fixp}{(\PTCp \toi \PTCp)}{\PTCp}$ for all the recursion
operators.
This operation is defines as in the case of the projective limit model
of \TC+\REC.
Theorem~\ref{thm-fix-nonexpansive} goes through for $\PTCp$.

The interpretation function $\Int{\ph}$ of terms with parameters from
$P$ is now defined by the equations given for $\Int{\ph}$ in
Section~\ref{sect-guarded-rec-eqns} and the following equation:
\begin{ldispl}
\Int{\pcc{t_1}{\NewThr{t_3}}{t_2}}(\rho) =
\pcc{\Int{t_1}(\rho)}{\NewThr{\Int{t_3}(\rho)}}{\Int{t_2}(\rho)}\;.
\end{ldispl}
The set $\GT{}{P}$ of guarded terms with parameters from $P$ is now
inductively defined by the rules given for $\GT{}{P}$ in
Section~\ref{sect-guarded-rec-eqns} and the following rule:
\begin{iteml}
\item
if $t_1,t_2,t_3 \in \PT{}{P}$, then
$\pcc{t_1}{\NewThr{t_3}}{t_2} \in \GT{}{P}$.
\end{iteml}
Lemmas~\ref{lemma-subst-value} and~\ref{lemma-guarded-term} and
Theorem~\ref{thm-unique-solution} go through for $\PTCp$.

Projective limit models of \TCf+\AIP\ and \TCf+\REC+\AIP\ are obtained
by expanding the projective limit models of \TCf\ and \TCf+\REC\ with
projection operations $\funct{\projop{n}}{\PTCp}{\PTCp}$.
These operations are defined as in the case of the projective limit
models of \TC+\AIP\ and \TC+\REC+\AIP.
Theorem~\ref{thm-proj-completeness}, Corollary~\ref{corol-decidability}
and Theorem~\ref{thm-co-rec-enum} go through for \TCf+\REC+\AIP.

It is easily proved that the projective limit model for \TC\ is a
submodel of the restriction of the projective limit model for \TCf\ to
the signature of \TC.

\section{Behavioural Equivalence of \PGLDf\ Programs}
\label{sect-PGLDf-eq}

In this short section, we introduce behavioural equivalence of \PGLDf\
programs and show that it is a $\rPi^0_1$-relation.

Let $\cP$ be the set of all \PGLDf\ programs.
Then, taking $\extr{\ph}$ as a function from $\cP$ to $\PTCp$, the
\emph{behavioural equivalence} relation ${\beqv} \subseteq \cP \x \cP$
is defined by $P \beqv Q$ iff $\extr{P}$ is identical to $\extr{Q}$ in
the projective limit model of \TCf+\REC+\AIP.

The following theorem is the counterpart of
Theorem~\ref{thm-co-rec-enum} in the world of \PGLDf\ programs.
\begin{theorem}
\label{thm-co-rec-enum-PGLDf}
The behavioural equivalence relation $\beqv$ is a $\rPi^0_1$-relation.
\end{theorem}
\begin{proof}
Let $\nm{Pr'} \subseteq \Nat \x \cP \x \cP$ be the relation defined by
$\nm{Pr'}(n,P,Q)$ iff $\proj{n}{\extr{P}}$ is identical to
$\proj{n}{\extr{Q}}$ in the projective limit model of \TC+\REC+\AIP.
By the definition of this model,
$P \beqv Q \Iff \Forall{n \in \Nat}{\nm{Pr'}(n,P,Q)}$
for all $P,Q \in \cP$.
Therefore, it is sufficient to prove that $\nm{Pr'}$ is a
$\rSigma^0_0$-relation.
In essentially the same way as described for \PGA\ programs
in~\cite{BL02a}, each \PGLDf\ program $P$ can be reduced to a \PGLDf\
program $Q$ without chains of jump instructions such that $\sextr{1}{P}$
is identical to $\sextr{1}{Q}$ in the projective limit model of
\TCf+\REC+\AIP.
This reduction is computable.
Moreover, each \PGLDf\ program $P$ without chains of jump instructions
can be translated into a closed term $p$ over the signature of
\TCf+\REC\ such that $\extr{P}$ is identical to the interpretation of
$p$ in the projective limit model of \TCf+\REC+\AIP.
Because it is restricted to programs without chains of jump
instructions, this translation is computable as well.
From this and Corollary~\ref{corol-decidability}, which goes through for
\TCf+\REC+\AIP, it follows that $\nm{Pr'}$ is a $\rSigma^0_0$-relation.
\qed
\end{proof}

\section{Conclusions}
\label{sect-concl}

In this paper, we have carried on the line of research with which we
made a start in~\cite{BM04c}.
We pursue with this line of research the object to develop a theory
about threads, multi-threading and interaction of threads with services
that is useful for (a)~gaining insight into the semantic issues
concerning the multi-threading related features found in contemporary
programming languages such as Java and C\#, and (b)~simplified formal
description and analysis of programs in which multi-threading is
involved.
In this paper, we have extended the theory developed in~\cite{BM04c}
with features that make it possible to deal with details of
multi-threading that come up where it is adjusted to the
object-orientation of those languages.
We regard this extension as just a step towards attaining the
above-mentioned object.
It is likely that applications of the theory developed so far will make
clear that multi-threading related features found in contemporary
programming languages are also intertwined with other matters and as a
consequence further developments are needed.

There is another line of research that emanated from the work presented
in~\cite{BM04c}.
That line of research concerns the development of a formal approach to
design new micro-architectures (architectures of micro-processors).
The approach should allow for the correctness of new micro-architectures
and their anticipated speed-up results to be verified.
In~\cite{BM06b}, we demonstrate the feasibility of an approach that
involves the use of thread algebra.
The line of research concerned is carried out in the framework of a
project investigating micro-threading~\cite{BJM96a,JL00a}, a technique
for speeding up instruction processing on a computer which requires that
programs are parallelized by judicious use of thread forking.

The work presented in this paper, was partly carried out in the
framework of that project as well.
For programs written in programming languages such as Java and C\#,
compilers will have to take care of the parallelization.
In~\cite{BM06d}, we investigate parallelization for simple programs,
which are close to machine language programs.
That work has convinced us that it is desirable to have available an
extension of thread algebra like the one presented in this paper when
developing parallelization techniques for the compilers referred to
above.

It is worth mentioning that the applications of thread algebra exceed
the domain of single non-distributed multi-threaded programs.
In~\cite{BM06a}, we extend the theory with features to cover systems
that consist of several multi-threaded programs on various hosts in
different networks.
To demonstrate its usefulness, we employ the extended theory to develop
a simplified, formal representation schema of the design of such systems
and to verify a property of all systems designed according to that
schema.
In~\cite{BM07a}, we extend the theory with features that allow for
details that come up with distributed multi-threading to be dealt with.
The features include explicit thread migration, load balancing and
capability searching.

\appendix

\section{Free and Bound Names, Substitution}
\label{app-fn-bn-subst}

In this appendix, we define $\FN^f(p)$, the set of free names of term
$p$ with respect to focus $f$, $\BN^f(p)$, the set of bound names of
term $p$ with respect to focus $f$, and $p \subst{s'}{s}^f$, the
substitution of name $s'$ for free occurrences of name $s$ with respect
to focus $f$ in term $p$.
In Table~\ref{eqns-fn-bn}, $\FN^f(p)$ and $\BN^f(p)$ are defined, and
in Table~\ref{eqns-subst}, $p \subst{s'}{s}^f$ is defined.
We write $m \subst{s'}{s}$, where $m \in \Meth$, for the result of
replacing in $m$ all occurrences of $s$ by $s'$.

\begin{table}[!tb]
\caption{Definition of $\FN^f(p)$ and $\BN^f(p)$}
\label{eqns-fn-bn}
\begin{eqntbl}
\begin{eqncol}
\FN^f(\Stop) = \emptyset                                              \\
\FN^f(\Dead) = \emptyset                                              \\
\FN^f(\Tau \bapf t) = \FN^f(t)                                        \\
\FN^f(\pcc{t}{g.m}{t'}) = \FN^f(t) \union \FN^f(t')
                                                 \qquad \mif f \neq g \\
\FN^f(\pcc{t}{f.m}{t'}) = \FN^f(t) \union \FN^f(t') \union \N(m)      \\
\FN^f(\csi{\alpha}) = \FN^f(\alpha)                                   \\
\FN^f(\std{t}) = \FN^f(t)                                             \\
\FN^f(\use{t}{g}{H}) = \FN^f(t)                                       \\
\FN^f(\local{g}{s}{t}) = \FN^f(t)                \hfill \mif f \neq g \\
\FN^f(\local{f}{s}{t}) = \FN^f(t) \diff \set{s}
\eqnsep
\FN^f(\emptyseq) = \emptyset                                          \\
\FN^f(\seq{t} \conc \alpha) = \FN^f(t) \union \FN^f(\alpha)
\end{eqncol}
\quad
\begin{eqncol}
\BN^f(\Stop) = \emptyset                                              \\
\BN^f(\Dead) = \emptyset                                              \\
\BN^f(\Tau \bapf t) = \BN^f(t)                                        \\
\BN^f(\pcc{t}{g.m}{t'}) = \BN^f(t) \union \BN^f(t')                   \\
\\
\BN^f(\csi{\alpha}) = \BN^f(\alpha)                                   \\
\BN^f(\std{t}) = \BN^f(t)                                             \\
\BN^f(\use{t}{g}{H}) = \BN^f(t)                                       \\
\BN^f(\local{g}{s}{t}) = \BN^f(t)                \qquad \mif f \neq g \\
\BN^f(\local{f}{s}{t}) = \BN^f(t) \union \set{s}
\eqnsep
\BN^f(\emptyseq) = \emptyset                                          \\
\BN^f(\seq{t} \conc \alpha) = \BN^f(t) \union \BN^f(\alpha)
\end{eqncol}
\end{eqntbl}
\end{table}

\begin{table}[!tb]
\caption{Definition of $p \subst{s'}{s}^f$}
\label{eqns-subst}
\begin{eqntbl}
\begin{eqncol}
\Stop \subst{s'}{s}^f = \Stop                                         \\
\Dead \subst{s'}{s}^f = \Dead                                         \\
(\Tau \bapf t) \subst{s'}{s}^f = \Tau \bapf (t \subst{s'}{s}^f)       \\
(\pcc{t}{g.m}{t'}) \subst{s'}{s}^f =
\pcc{(t \subst{s'}{s}^f)}{g.m}{(t' \subst{s'}{s}^f)}
                                                 \hfill \mif f \neq g \\
(\pcc{t}{f.m}{t'}) \subst{s'}{s}^f =
\pcc{(t \subst{s'}{s}^f)}{f.m \subst{s'}{s}}{(t' \subst{s'}{s}^f)}    \\
\csi{\alpha} \subst{s'}{s}^f = \csi{\alpha \subst{s'}{s}^f}           \\
\std{t} \subst{s'}{s}^f = \std{t \subst{s'}{s}^f}                     \\
(\use{t}{g}{H}) \subst{s'}{s}^f = \use{(t \subst{s'}{s}^f)}{g}{H}     \\
\local{g}{s''}{t} \subst{s'}{s}^f = \local{g}{s''}{t \subst{s'}{s}^f}
             \hfill \mif f = g \Implies (s \neq s'' \And s' \neq s'')
\quad \\
\local{f}{s''}{t} \subst{s'}{s}^f =
\local{f}{s'''}{(t \subst{s'''}{s''}^f) \subst{s'}{s}^f}
                               \qquad \mif (s \neq s'' \And s' = s'')
\phantom{f = g \Implies {}} \quad \\
     \hfill (s''' \not\in \FN^f(t) \union \BN^f(t) \union \set{s,s'}) \\
\local{f}{s}{t} \subst{s'}{s}^f = \local{f}{s}{t }
\eqnsep
\emptyseq \subst{s'}{s}^f = \emptyseq                                 \\
(\seq{t} \conc \alpha) \subst{s'}{s}^f =
\seq{t \subst{s'}{s}^f} \conc (\alpha \subst{s'}{s}^f)
\end{eqncol}
\end{eqntbl}
\end{table}

\section{CPO Structure for Projective Limit Model}
\label{app-projlim-cpo}

In this appendix, we make $\PTC$ into a complete partial ordering (cpo)
to establish the existence of least solutions of recursion equations
using Tarski's fixed point theorem.

The \emph{approximation} relation ${\apx} \subseteq \ITC \x \ITC$ is
the smallest partial ordering such that for all $p,p',q,q' \in \ITC$:
\begin{iteml}
\item
$\Dead \apx p$;
\item
$p \apx p' \Implies
 \Tau \bapf p \apx \Tau \bapf p'$;
\item
for all $f \in \Foci$ and $m \in \Meth$,
$p \apx p' \And q \apx q' \Implies
 \pcc{p}{f.m}{q} \apx \pcc{p'}{f.m}{q'}$;
\item
for all $f \in \Foci$, $m \in \Meth$, and $s_1,\ldots,s_n \in \N(m)$
with $s_i \neq s_j$ for all $i,j \in [1,n]$ for which $i \neq j$,
$p \apx p' \And q \apx q' \Implies
 \smash{\local{f}{s_1}{\ldots\local{f}{s_n}{\pcc{p}{f.m}{q}}\ldots}}
  \apx
 \smash{\local{f}{s_1}{\ldots\local{f}{s_n}{\pcc{p'}{f.m}{q'}}\ldots}}$.
\end{iteml}
The \emph{approximation} relation ${\apx} \subseteq \PTC \x \PTC$ is
defined component-wise:%
\begin{ldispl}
\projseq{p_n}{n} \apx \projseq{q_n}{n} \Iff
\Forall{n \in \Nat}{p_n \apx q_n}\;.
\end{ldispl}
The approximation relation $\apx$ on $\ITCn{n}$ is simply the
restriction of $\apx$ on $\ITC$ to $\ITCn{n}$.

The following proposition states that any $p \in \ITC$ is finitely
approximated by projection.
\begin{proposition}
\label{prop-fin-apx}
For all $p \in \ITC$:
\begin{ldispl}
\Exists{n \in \Nat}
 {(\Forall{k < n}{\proj{k}{p} \apx \proj{k+1}{p}} \And
   \Forall{l \geq n}{\proj{l}{p} = p})}\;.
\end{ldispl}
\end{proposition}
\begin{proof}
The proof follows the same line as the proof of Proposition~1
from~\cite{BB03a}.
This means that it is a rather trivial proof by induction on the
structure of $p$.
Here, we have to consider the additional case
$p \equiv
 \local{f}{s_1}{\ldots\local{f}{s_n}{\pcc{p'}{f.m}{p''}}\ldots}$
with $s_1,\ldots,s_n \in \N(m)$ and $s_i \neq s_j$ for all
$i,j \in [1,n]$ for which $i \neq j$.
This case goes analogous to the case $p \equiv \pcc{p'}{f.m}{p''}$.
\qed
\end{proof}

The properties stated in the following lemma will be used in the proof
of Theorem~\ref{thm-projlim-cpo} given below.
\begin{lemma}
\label{lemma-proj-system}
For all $n \in \Nat$:
\begin{enuml}
\item
\label{fin-cpo}
$\tup{\ITCn{n},\apx}$ is a cpo;
\item
\label{pin-cont}
$\projop{n}$ is continuous;
\item
\label{pin-props}
for all $p \in \ITC$:
\begin{enuml}
\item
\label{pin-props-1}
$\proj{n}{p} \apx p$;
\item
\label{pin-props-2}
$\proj{n}{\proj{n}{p}} = \proj{n}{p}$;
\item
\label{pin-props-3}
$\proj{n+1}{\proj{n}{p}} = \proj{n}{p}$.
\end{enuml}
\end{enuml}
\end{lemma}
\begin{proof}
The proof follows similar lines as the proof of Proposition~2
from~\cite{BB03a}.
For property~\ref{fin-cpo}, we now have to consider directed sets that
consist of $\Dead$, postconditional compositions and restrictions of
postconditional compositions instead of $\Dead$ and postconditional
compositions.
However, the same reasoning applies.
For property~\ref{pin-cont}, we now have to use induction on the
structure of the elements of $\ITC$ and distinction between the cases
$n = 0$ and $n > 0$ for postconditional compositions.
Due to the presence of restrictions, we cannot use induction on $n$ and
case distinction on the structure of the elements of $\ITC$
like in~\cite{BB03a}.
However, the crucial details of the proof remain the same.
Like in~\cite{BB03a}, property~\ref{pin-props-1} follows immediately
from Proposition~\ref{prop-fin-apx}.
Properties~\ref{pin-props-2} and~\ref{pin-props-3} follow immediately
from Lemma~\ref{lemma-min-proj}.
\qed
\end{proof}

The following theorem states some basic properties of the approximation
relation $\apx$ on $\PTC$.
\begin{theorem}
\label{thm-projlim-cpo}
$\tup{\PTC,\apx}$ is a cpo with
$\lub P = \projseq{\lub \set{\proj{n}{p} \where p \in P}}{n}$ for all
directed sets $P \subseteq \PTC$.
Moreover, up to (order) isomorphism $\ITC \subseteq \PTC$.
\end{theorem}
\begin{proof}
The proof follows the same line as the proof of Theorem~1
from~\cite{BB03a}.
That is, using general properties of the projective limit construction
on cpos, the first part follows immediately from
Lemmas~\ref{lemma-proj-system}.\ref{fin-cpo}
and~\ref{lemma-proj-system}.\ref{pin-cont},
and the second part follows easily from Proposition~\ref{prop-fin-apx}
and Lemma~\ref{lemma-proj-system}.\ref{pin-props}.
\qed
\end{proof}

Another important property of the approximation relation $\apx$ on
$\PTC$ is stated in the following theorem.
\begin{theorem}
\label{thm-projlim-cont}
The operations from the projective limit model of \TC\ are continuous
with respect to $\apx$.
\end{theorem}
\begin{proof}
The proof begins by establishing the monotonicity of the operations on
$\ITC$.
For the postconditional composition operations, this follows
immediately from the definition of $\apx$ on $\ITC$.
For the cyclic interleaving operation, it is straightforwardly proved by
induction on the sum of the depths plus one of the threads in the thread
vector and case distinction on the structure of the first thread in the
thread vector.
For the deadlock at termination operation, the thread-service
composition operations and the restriction operations, it is easily
proved by structural induction.
Then the monotonicity of the operations on $\PTC$ follows from their
monotonicity on $\ITC$, the monotonicity of the projection operations
and the definition of $\apx$ on $\PTC$.

What remains to be proved is that least upper bounds of directed sets
are preserved by the operations.
We will show how the proof goes for the postconditional composition
operations.
The proofs for the other kinds of operations go similarly.
Let $P,Q \subseteq \PTC$ be directed sets.
Then, for all $n \in \Nat$,
$\set{\proj{n}{p} \where p \in P},
 \set{\proj{n}{q} \where q \in Q},
 \set{\pcc{\proj{n}{p}}{a}{\proj{n}{q}} \where
              p \in P \And q \in Q} \subseteq \ITCn{n}$
are directed sets by the monotonicity of $\projop{n}$.
Moreover, it is easily proved by induction on $n$, using the definition
of $\apx$ on $\ITCn{n}$, that these directed sets are finite.
This implies that they have maximal elements.
From this, it follows by the monotonicity of $\pcc{\ph}{a}{\ph}$ that,
for all $n \in \Nat$,
$\pcc{(\lub \set{\proj{n}{p} \where p \in P})}{a}
     {(\lub \set{\proj{n}{q} \where q \in Q})} =
 \lub \set{\pcc{\proj{n}{p}}{a}{\proj{n}{q}} \where
           p \in P \And q \in Q}$.
From this, it follows by the property of lubs of directed sets stated in
Theorem~\ref{thm-projlim-cpo} and the definition of $\projop{n+1}$ that,
for all $n \in \Nat$,
$\proj{n+1}{\pcc{(\lub P)}{a}{(\lub Q)}} =
 \proj{n+1}{\lub \set{\pcc{p}{a}{q} \where p \in P \And q \in Q}}$.
Because
$\proj{0}{\pcc{(\lub P)}{a}{(\lub Q)}} = \Dead =
 \proj{0}{\lub \set{\pcc{p}{a}{q} \where p \in P \And q \in Q}}$,
also for all $n \in \Nat$,
$\proj{n}{\pcc{(\lub P)}{a}{(\lub Q)}} =
 \proj{n}{\lub \set{\pcc{p}{a}{q} \where p \in P \And q \in Q}}$.
From this, it follows by the definition of $\apx$ on $\PTC$ that
$\pcc{(\lub P)}{a}{(\lub Q)} =
 \lub \set{\pcc{p}{a}{q} \where p \in P \And q \in Q}$.
\qed
\end{proof}

We have the following result about fixed points.
\begin{theorem}
\label{thm-least-solution}
Let $x$ be a variable, and
let $t$ be a term over the signature of \TC\ in which no other variables
than $x$ have free occurrences.
Then $\Int{t}^x$ has a least fixed point with respect to $\apx$, i.e.\
there exists a  $p \in \PTC$ such that $\Int{t}^x(p) = p$ and, for all
$q \in \PTC$, $\Int{t}^x(q) = q$ implies $p \apx q$.
\end{theorem}
\begin{proof}
We have from Theorem~\ref{thm-projlim-cpo} that $\tup{\PTC,\apx}$ is a
cpo and, using Theorem~\ref{thm-projlim-cont}, it is easily proved by
induction on the structure of $t$ that $\Int{t}^x$ is continuous.
From this, we conclude by Tarski's fixed point theorem that $\Int{t}^x$
has a least fixed point with respect to $\apx$.
\qed
\end{proof}
Hence, every recursion equation in which no recursion operator occurs
has a least solution in the projective limit model for \TC.

According to Tarski's fixed point theorem, the least fixed point of a
continuous operation $\funct{\phi}{\PTC}{\PTC}$ is
$\lub \set{\phi^n(\Dead) \where n \in \Nat}$.
It is well-known that the restriction to continuous functions of the
operation $\funct{\lfixp}{(\PTC \to \PTC)}{\PTC}$ defined by
$\lfixp(\phi) = \lub \set{\phi^n(\Dead) \where n \in \Nat}$ is
continuous.
Moreover, for all functions $\funct{\phi}{\PTC}{\PTC}$ that are
representable by a term over the signature of \TC+\REC,
$\fixp(\phi) = \lfixp(\phi)$.
This brings us to the following corollary of
Theorem~\ref{thm-least-solution}.
\begin{corollary}
\label{corol-least-solution}
Let $x$ be a variable, and let $t$ be a term over the signature of
\TC\textup{+}\REC\ in which no other variables than $x$ have free
occurrences.
Then $\Int{t}^x$ has a least fixed point with respect to $\apx$.
\end{corollary}
Hence, every recursion equation has a least solution in the projective
limit model for \TC\textup{+}\REC.

\bibliographystyle{spmpsci}
\bibliography{TA}

\begin{thebibliography}{10}
\providecommand{\url}[1]{{#1}}
\providecommand{\urlprefix}{URL }
\expandafter\ifx\csname urlstyle\endcsname\relax
  \providecommand{\doi}[1]{DOI~\discretionary{}{}{}#1}\else
  \providecommand{\doi}{DOI~\discretionary{}{}{}\begingroup
  \urlstyle{rm}\Url}\fi

\bibitem{AFV00a}
Aceto, L., Fokkink, W.J., Verhoef, C.: Structural operational semantics.
\newblock In: J.A. Bergstra, A.~Ponse, S.A. Smolka (eds.) Handbook of Process
  Algebra, pp. 197--292. Elsevier, Amsterdam (2001)

\bibitem{AMST97a}
Agha, G.A., Mason, I.A., Smith, S.F., Talcott, C.L.: A foundation for actor
  computation.
\newblock Journal of Functional Programming \textbf{7}(1), 1--72 (1997)

\bibitem{BBKM84a}
de~Bakker, J.W., Bergstra, J.A., Klop, J.W., Meyer, J.J.C.: Linear time and
  branching time semantics for recursion with merge.
\newblock Theoretical Computer Science \textbf{34}, 135--156 (1984)

\bibitem{BZ82a}
de~Bakker, J.W., Zucker, J.I.: Processes and the denotational semantics of
  concurrency.
\newblock Information and Control \textbf{54}(1/2), 70--120 (1982)

\bibitem{BB02a}
Bergstra, J.A., Bethke, I.: Molecular dynamics.
\newblock Journal of Logic and Algebraic Programming \textbf{51}, 193--214
  (2002)

\bibitem{BB03a}
Bergstra, J.A., Bethke, I.: Polarized process algebra and program equivalence.
\newblock In: J.C.M. Baeten, J.K. Lenstra, J.~Parrow, G.J. Woeginger (eds.)
  Proceedings 30th ICALP, \emph{Lecture Notes in Computer Science}, vol. 2719,
  pp. 1--21. Springer-Verlag (2003)

\bibitem{BK82a}
Bergstra, J.A., Klop, J.W.: Fixed point semantics in process algebra.
\newblock CWI Report IW~206/82, Centre for Mathematics and Computer Science
  (1982)

\bibitem{BK84b}
Bergstra, J.A., Klop, J.W.: Process algebra for synchronous communication.
\newblock Information and Control \textbf{60}(1/3), 109--137 (1984)

\bibitem{BL02a}
Bergstra, J.A., Loots, M.E.: Program algebra for sequential code.
\newblock Journal of Logic and Algebraic Programming \textbf{51}(2), 125--156
  (2002)

\bibitem{BM05a}
Bergstra, J.A., Middelburg, C.A.: Splitting bisimulations and retrospective
  conditions.
\newblock Information and Computation \textbf{204}(7), 1083--1138 (2006)

\bibitem{BM05c}
Bergstra, J.A., Middelburg, C.A.: Thread algebra with multi-level strategies.
\newblock Fundamenta Informaticae \textbf{71}(2/3), 153--182 (2006)

\bibitem{BM06d}
Bergstra, J.A., Middelburg, C.A.: Synchronous cooperation for explicit
  multi-threading.
\newblock Acta Informatica \textbf{44}(7/8), 525--569 (2007)

\bibitem{BM04c}
Bergstra, J.A., Middelburg, C.A.: Thread algebra for strategic interleaving.
\newblock Formal Aspects of Computing \textbf{19}(4), 445--474 (2007)

\bibitem{BM06a}
Bergstra, J.A., Middelburg, C.A.: A thread algebra with multi-level strategic
  interleaving.
\newblock Theory of Computing Systems \textbf{41}(1), 3--32 (2007)

\bibitem{BM08d}
Bergstra, J.A., Middelburg, C.A.: Data linkage algebra, data linkage dynamics,
  and priority rewriting.
\newblock Electronic Report PRG0806, Programming Research Group, University of
  Amsterdam (2008).
\newblock Available from {\tt
  http://www.science.uva.nl/\linebreak[2]research/prog/publications.html}.
  Also available from {\tt http://arxiv.org/}: {\tt arXiv:0804.4565v2 [cs.LO]}

\bibitem{BM07a}
Bergstra, J.A., Middelburg, C.A.: Distributed strategic interleaving with load
  balancing.
\newblock Future Generation Computer Systems \textbf{24}(6), 530--548 (2008)

\bibitem{BM06b}
Bergstra, J.A., Middelburg, C.A.: {Maurer} computers for pipelined instruction
  processing.
\newblock Mathematical Structures in Computer Science \textbf{18}(2), 373--409
  (2008)

\bibitem{BP02a}
Bergstra, J.A., Ponse, A.: Combining programs and state machines.
\newblock Journal of Logic and Algebraic Programming \textbf{51}(2), 175--192
  (2002)

\bibitem{BB92d}
Berry, G., Boudol, G.: The chemical abstract machine.
\newblock Theoretical Computer Science \textbf{96}, 217--248 (1992)

\bibitem{BJM96a}
Bolychevsky, A., Jesshope, C.R., Muchnick, V.: Dynamic scheduling in {RISC}
  architectures.
\newblock IEE Proceedings Computers and Digital Techniques \textbf{143}(5),
  309--317 (1996)

\bibitem{Cro89a}
Croom, F.H.: Principles of Topology.
\newblock Saunders College Publishing, Philadelphia (1989)

\bibitem{Dug66a}
Dugundji, J.: Topology.
\newblock Allyn and Bacon, Boston (1966)

\bibitem{GJSB00a}
Gosling, J., Joy, B., Steele, G., Bracha, G.: The {Java} Language
  Specification, second edn.
\newblock Addison-Wesley, Reading, MA (2000)

\bibitem{HWG03a}
Hejlsberg, A., Wiltamuth, S., Golde, P.: {C\#} Language Specification.
\newblock Addison-Wesley, Reading, MA (2003)

\bibitem{Hod93a}
Hodges, W.A.: Model Theory, \emph{Encyclopedia of Mathematics and Its
  Applications}, vol.~42.
\newblock Cambridge University Press, Cambridge (1993)

\bibitem{JL00a}
Jesshope, C.R., Luo, B.: Micro-threading: A new approach to future {RISC}.
\newblock In: Australian Computer Architecture Conference 2000, pp. 34--41.
  IEEE Computer Society Press (2000)

\bibitem{Kra87a}
Kranakis, E.: Fixed point equations with parameters in the projective model.
\newblock Information and Computation \textbf{75}(3), 264--288 (1987)

\bibitem{Men97a}
Mendelson, E.: Introduction to Mathematical Logic, fourth edn.
\newblock Chapman and Hall, London (1997)

\bibitem{Mid01a}
Middelburg, C.A.: An alternative formulation of operational conservativity with
  binding terms.
\newblock Journal of Logic and Algebraic Programming \textbf{55}(1/2), 1--19
  (2003)

\bibitem{MPW92c}
Milner, R., Parrow, J., Walker, D.: A calculus of mobile processes.
\newblock Information and Computation \textbf{100}, 1--77 (1992)

\bibitem{MGR05a}
Mousavi, M.R., Gabbay, M.J., Reniers, M.A.: {SOS} for higher order processes.
\newblock In: M.~Abadi, L.~de~Alfaro (eds.) CONCUR 2005, \emph{Lecture Notes in
  Computer Science}, vol. 3653, pp. 308--322. Springer-Verlag (2005)

\bibitem{PU01a}
Ponse, A., Usenko, Y.S.: Equivalence of recursive specifications in process
  algebra.
\newblock Information Processing Letters \textbf{80}, 59--65 (2001)

\bibitem{ST99a}
Sannella, D., Tarlecki, A.: Algebraic preliminaries.
\newblock In: E.~Astesiano, H.J. Kreowski, B.~Krieg-Br{\"{u}}ckner (eds.)
  Algebraic Foundations of Systems Specification, pp. 13--30. Springer-Verlag,
  Berlin (1999)

\bibitem{Sch86a}
Schmidt, D.A.: Denotational Semantics: A Methodology for Language Development.
\newblock Allyn and Bacon, Boston (1986)

\bibitem{Sho91a}
Shoenfield, J.R.: Recursion Theory.
\newblock Lecture Notes in Logic. Springer-Verlag, Berlin (1991)

\bibitem{ST91a}
Stoltenberg-Hansen, V., Tucker, J.V.: Algebraic and fixed point equations over
  inverse limits of algebras.
\newblock Theoretical Computer Science \textbf{87}, 1--24 (1991)

\bibitem{Wir90a}
Wirsing, M.: Algebraic specification.
\newblock In: J.~van Leeuwen (ed.) Handbook of Theoretical Computer Science,
  vol.~B, pp. 675--788. Elsevier, Amsterdam (1990)

\bibitem{Wir71a}
Wirth, N.: The programming language {PASCAL}.
\newblock Acta Informatica \textbf{1}(1), 35--63 (1971)

\end{thebibliography}

\end{document}